\input harvmac.tex
\input epsf.tex
\parindent=0pt
\parskip=5pt

\hyphenation{satisfying}

\def\IR{{\hbox{{\rm I}\kern-.2em\hbox{\rm R}}}}
\def\IB{{\hbox{{\rm I}\kern-.2em\hbox{\rm B}}}}
\def\IN{{\hbox{{\rm I}\kern-.2em\hbox{\rm N}}}}
\def\IC{\,\,{\hbox{{\rm I}\kern-.59em\hbox{\bf C}}}}
\def\IZ{{\hbox{{\rm Z}\kern-.4em\hbox{\rm Z}}}}
\def\IP{{\hbox{{\rm I}\kern-.2em\hbox{\rm P}}}}
\def\IH{{\hbox{{\rm I}\kern-.4em\hbox{\rm H}}}}
\def\ID{{\hbox{{\rm I}\kern-.2em\hbox{\rm D}}}}
\def\II{{\hbox{\rm I}\kern-.2em\hbox{\rm I}}}
\def\mmu{{\mu}\hskip-.55em{\mu}}
\def\zzeta{{\zeta}\hskip-.45em{\zeta}}

\noblackbox

\leftline{\epsfxsize1.0in\epsfbox{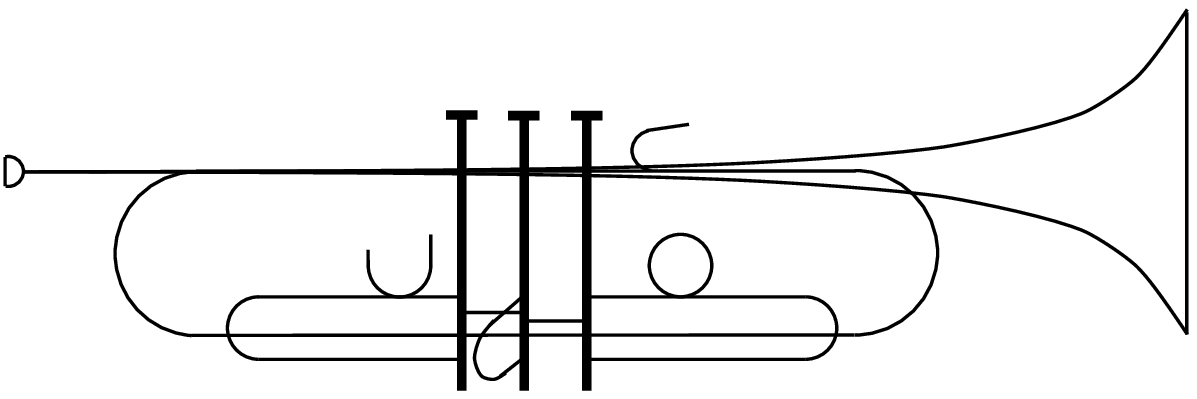}}
\vskip-0.9cm
\Title{\vbox{\baselineskip12pt
\hbox{UK/97--22}
\hbox{hep-th/9711082}}}
{Anatomy of a Duality}

\centerline{\bf Clifford V. Johnson$^\dagger$}

\bigskip
\bigskip

\vbox{\baselineskip12pt\centerline{\hbox{\it Department of Physics and 
Astronomy}}
\centerline{\hbox{\it University of Kentucky}}
\centerline{\hbox{\it Lexington, KY 40506--0055 USA}}}
\footnote{}{\sl email: $^\dagger${\tt cvj@pa.uky.edu}}
\vskip1.7cm
\centerline{\bf Abstract}
\vskip1.7cm
\vbox{\narrower\baselineskip=12pt\noindent

The nature of M--theory on $K3{\times}{\cal I}$, where $\cal I$ is a
line interval, is considered, with a view towards formulating a
``matrix theory'' representation of that situation.  Various limits of
this compactification of M--theory yield a number of well known ${\cal
N}{=}1$ six dimensional compactifications of the heterotic and type~I
string theories. Geometrical relations between these limits give rise
to string/string dualities between some of these compactifications.  At
a special point in the moduli space of compactifications, this
motivates a partial definition of the matrix theory representation of
the M--theory on $K3{\times}{\cal I}$ as the large $N$ limit of a
certain type~IA orientifold model probed by a conglomerate of $N$
D--branes. Such a definition in terms of D--branes and orientifold
planes is suggestive, but necessarily incomplete, due to the low
amount of superymmetry. It is proposed ---following hints from the
orientifold model--- that the complete matrix theory representation of
the $K3{\times}{\cal I}$ compactified M--theory is given by the large
$N$ limit of  compactification ---on a suitable ``dual'' surface---
of the ``little  heterotic string'' ${\cal N}{=}1$ six
dimensional quantum theories.}


\Date{11th November 1997}
\baselineskip13pt
\lref\dbranes{J.~Dai, R.~G.~Leigh and J.~Polchinski, 
{\sl `New Connections Between String Theories'}, Mod.~Phys.~Lett.
{\bf A4} (1989) 2073\semi P.~Ho\u{r}ava, {\sl `Background Duality of
Open String Models'}, Phys. Lett. {\bf B231} (1989) 251\semi
R.~G.~Leigh, {\sl `Dirac--Born--Infeld Action from Dirichlet Sigma
Model'}, Mod.~Phys.~Lett. {\bf A4} (1989) 2767\semi J.~Polchinski,
{\sl `Combinatorics Of Boundaries in String Theory'}, Phys.~Rev.~{\bf D50}
(1994) 6041, hep-th/9407031.}

\lref\orientifolds{A. Sagnotti, in {\sl `Non--Perturbative Quantum
 Field Theory'}, Eds. G. Mack {\it et. al.} (Pergammon Press, 1988), p.521\semi
V. Periwal, unpublished\semi J. Govaerts, Phys. Lett. {\bf B220}
(1989) 77\semi P. Hor\u{a}va, {\sl `Strings on World Sheet Orbifolds'}, 
Nucl. Phys. {\bf B327} (1989) 461.}
\lref\nsfivebrane{S.-J. Rey, {\sl `Axionic String Instantons and Their 
Low--Energy Implications'}, in {\sl `Superstrings and Particle
Theory'}, proceedings, eds.  L. Clavelli and B. Harms, World
Scientific, 1990\semi S.-J. Rey, {\sl `The Confining Phase of
Superstrings and Axionic Strings'}, Phys. Rev. {\bf D43} {\bf 1991}
526\semi A. Strominger, {\sl `Heterotic Solitons'} Nucl. Phys.  {\bf
B343}, (1990) 167; {\it Erratum: ibid.}, {\bf 353} (1991) 565.}

\lref\mtwo{E. Bergshoeff, E. Sezgin and P. K. Townsend,
 {\sl `Supermembranes and Eleven Dimensional Supergravity'}, Phys.
Lett. {\bf B189} (1987) 75\semi M. J. Duff and K. S. Stelle, {\sl
`Multimembrane solutions of D=11 Supergravity'}, Phys. Lett. {\bf
B253} (1991) 113.}
\lref\mfive{R. G\"uven, {\sl `Black $p$--Brane 
Solutions of D=11 Supergravity theory'}, Phys. Lett. {\bf B276} (1992)
49.}

\lref\town{P. Townsend, {\sl `The eleven-dimensional supermembrane revisited'},
  Phys. Lett. {\bf B350} (1995) 184, hep-th/9501068}
\lref\goed{E. Witten, {\sl `String Theory Dynamics in Various Dimensions'}, 
Nucl. Phys. {\bf B443} (1995) 85, hep-th/9503124.}

\lref\romans{L. Romans, {\sl `Massive N=2a Supergravity In Ten-Dimensions'},
 Phys. Lett. {\bf B169} (1986) 374.}
\lref\others{E. Bergshoeff, M. de Roo, M. B. Green, G. Papadopoulos
 and P. K. Townsend, 
{\sl `Duality of Type II 7-branes and 8-branes'}, Nucl.Phys. {\bf
B470} (1996) 113, hep-th/9601150.}

\lref\duff{Duff, Minasian and Witten, {\sl `Evidence for Heterotic/Heterotic 
Duality'}, Nucl. Phys. {\bf B465} (1996) 413, hep-th/9601036.}
\lref\duffetal{M. J. Duff, {\sl 
`Strong/Weak Coupling Duality from the Dual String', Nucl. Phys. {\bf B442}
(1995) 47, hep-th/9501030.}\semi M. J. Duff, {\sl `Putting
string/string duality to the test', Nucl. Phys. {\bf B436} (1995) 507,
hep-th/9406198.}}
\lref\berkoozi{M. Berkooz, R. G. Leigh, J. Polchinski, J. Schwarz, N. Seiberg 
and E. Witten, {\sl `Anomalies, Dualities, and Topology of D=6 N=1 Superstring
Vacua'}, Nucl. Phys. {\bf B475} (1996) 115, hep-th/9605184.}
\lref\rozali{M. Rozali,
 {\sl `Matrix Theory and U--Duality in Seven Dimensions'}, 
Phys. Lett. {\bf B400} (1997) 260, hep-th/9702136.}
\lref\berkoozii{M. Berkooz, N. Seiberg and M. Rozali, {\sl 
`Matrix Description of 
M-theory on $T^4$ and $T^5$'}, Phys. Lett. {\bf B408}
 (1997) 105, hep-th/9704089.}
\lref\berkooziii{M. Berkooz, and M. Rozali, 
{\sl `String Dualities from Matrix Theory'}, hep-th/9705175.}

\lref\atish{A. Dahbolkar and J. Park, {\sl `Strings on Orientifolds'},
 Nucl. Phys.  {\bf B477} (1996) 701, hep-th/9604178.}
\lref\bfss{T. Banks, W. Fischler, S. Shenker and L. Susskind, {\sl 
 `M--Theory As A 
Matrix Model: A Conjecture'}, Phys. Rev. {\bf D55}
 (1997) 5112, hep-th/9610043.}
\lref\edjoe{J. Polchinksi and E. Witten,  {\sl 
`Evidence for Heterotic - Type I 
String Duality'}, Nucl. Phys. {\bf B460} (1996) 525, hep-th/9510169.}
\lref\seiberg{N. Seiberg, {\sl `Matrix Description of M-theory on $T^5$ and 
$T^5/Z_2$'}, Phys. Lett. {\bf B408} (1997) 98, hep-th/9705221.}
\lref\sagnotti{M. Bianchi and A. Sagnotti, {\sl `Twist
 Symmetry and Open String Wilson Lines'} Nucl. Phys. {\bf B361} (1991)
519.}
\lref\sagnottii{A. Sagnotti, {\sl 
`A Note on the Green - Schwarz Mechanism in Open - String Theories'},
Phys. Lett. {\bf B294} (1992) 196, hep-th/9210127.}
\lref\horavawitten{P. Horava and E. Witten, {\sl `Heterotic and Type I String 
Dynamics from Eleven Dimensions'}, Nucl. Phys. {\bf B460} (1996) 506,
hep-th/9510209.}

\lref\dvv{R. Dijkgraaf, E. Verlinde, H. Verlinde, 
{\sl `Matrix String Theory'}, Nucl. Phys. {\bf B500} (1997) 43,
hep-th/9703030.}
\lref\dvvii{R. Dijkgraaf, E. Verlinde, H. Verlinde, {\sl `5D Black Holes and 
Matrix Strings'}, hep-th/9704018.}
\lref\robme{C. V. Johnson and R. C. Myers, 
{\sl `Aspects of Type IIB Theory on Asymptotically Locally Euclidean
Spaces'}, Phys. Rev. {\bf D55} (1997) 6382, hep-th/9610140.}
\lref\douglasi{M. R. Douglas, {\sl `Branes within Branes'}, hep-th/9512077.}

\lref\mumford{D. Mumford and J. Fogarty, {\sl `Geometric Invariant Theory'},
Springer, 1982.}
\lref\edcomm{E. Witten, {\sl `Some Comments On String Dynamics'}, in the 
Proceedings of {\sl Strings 95}, USC, 1995, hep-th/9507121.}

\lref\joetensor{J. Polchinski, {\sl `Tensors From $K3$ Orientifolds'}, 
hep-th/9606165.}
\lref\kronheimer{P. B. Kronheimer, {\sl `The Construction of ALE Spaces as 
Hyper--K\"ahler Quotients'}, J.~Diff. Geom. {\bf 29} (1989) 665.}
\lref\hitchin{N. J.  Hitchin,  {\sl `Polygons and Gravitons'}, Math. Proc. 
Camb. Phil. Soc. {\bf 85} (1979) 465.}
\lref\douglasmoore{M. R. Douglas and G. Moore,
  {\sl `D--Branes, Quivers and ALE Instantons'}, hep-th/9603167.}

\lref\hitchinetal{N. J. Hitchin, A. Karlhede, U. Lindstr\"om and M. Ro\u cek, 
{\sl `Hyper--K\"ahler Metrics and Supersymmetry'}, Comm. Math. Phys. {\bf 108}
(1987) 535.}

\lref\klein{F. Klein, {\sl `Vorlesungen \"Uber das Ikosaeder und die 
Aufl\"osung der Gleichungen vom f\"unften Grade'}, Teubner, Leipzig 1884;
F. Klein, {\sl `Lectures on the Icosahedron and  the Solution of an Equation 
of Fifth Degree'}, Dover, New York, 1913.}

\lref\elliot{J. P. Elliot and P. G. Dawber, {\sl `Symmetry in Physics'}, 
McMillan, 1986.}

\lref\ericjoe{E. G. Gimon and J. Polchinski, {\sl `Consistency
 Conditions of Orientifolds and D--Manifolds'}, Phys. Rev. {\bf D54} (1996) 
1667, hep-th/9601038.}
\lref\ericmeI{E. G. Gimon and C. V. Johnson, {\sl `$K3$ Orientifolds'}, Nucl. 
Phys. {\bf B477} (1996) 715, hep-th/9604129.}
\lref\ericmeII{E. G. Gimon and C. V. Johnson, {\sl `Multiple Realisations of 
${\cal N}{=}1$ Vacua in Six Dimensions'}, Nucl. Phys. {\bf B479}
(1996), 285, hep-th/9606176}
\lref\mackay{J. McKay, {\sl `Graphs, Singularties
 and Finite Groups'}, Proc. Symp. Pure. Math. {\bf 37} (1980) 183,
Providence, RI; Amer. Math. Soc.}

\lref\orbifold{L. Dixon, J. Harvey, C. Vafa and E. Witten, {\sl `Strings on 
Orbifolds'}, Nucl. Phys. {\bf B261} (1985) 678;
{\it ibid}, Nucl. Phys. {\bf B274} (1986) 285.}
\lref\algebra{P. Slodowy, {\sl `Simple Singularities and Simple Algebraic 
Groups'}, Lecture Notes in Math., Vol.  {\bf 815}, Springer, Berlin, 1980.}
\lref\gibhawk{G. W. Gibbons and S. W. Hawking, {\sl `Gravitational
Multi--Instantons'}, Phys. Lett. {\bf B78} (1978) 430.}
\lref\eguchihanson{T. Eguchi and A. J. Hanson, {\sl `Asymptotically Flat 
Self--Dual Solutions to Euclidean Gravity'}, Phys. Lett. {\bf B74} (1978) 249.}

\lref\wittenadhm{E. Witten, {\sl `Sigma Models and the ADHM Construction of 
Instantons'}, J.~Geom.  Phys. {\bf 15} (1995) 215, hep-th/9410052.}
\lref\edsmall{E. Witten, {\sl `Small Instantons in String Theory'},  Nucl. 
Phys. {\bf B460} (1996) 541, hep-th/9511030.}
\lref\douglasii{M. R.  Douglas, {\sl `Gauge Fields and D--Branes'},  
hep-th/9604198.}

\lref\phases{E. Witten, {\sl `Phases of $N{=}2$ Theories in Two Dimensions'}, 
Nucl. Phys. {\bf B403} (1993) 159,  hep-th/9301042.}
\lref\edbound{E. Witten, {\sl `Bound States of Strings and $p$--Branes'}, 
Nucl. Phys. {\bf B460} (1996) 335, hep-th/9510135.}

\lref\ADHM{M. F. Atiyah, V. Drinfeld, N. J. Hitchin and Y. I. Manin, {\sl
`Construction of Instantons'} Phys. Lett. {\bf A65} (1978) 185.}

\lref\kronheimernakajima{P. B. Kronheimer and H. Nakajima, {\sl `Yang--Mills 
Instantons on ALE Gravitational Instantons'}, Math. Ann. {\bf 288} (1990) 263.}
\lref\italiansi{M. Bianchi, F. Fucito, G. Rossi, and M. Martinelli, 
{\sl `Explicit Construction of Yang--Mills Instantons on ALE Spaces'}, 
Nucl. Phys. {\bf B473} (1996) 367, hep-th/9601162.}
\lref\italiansii{D. Anselmi, M. Bill\'o, P. Fr\'e, L. Giraradello and A. 
Zaffaroni, {\sl `ALE Manifolds and Conformal Field Theories'},  Int. J. 
Mod. Phys. {\bf A9} (1994) 3007,  hep-th/9304135.}

\lref\nonrenorm{L. Alvarez--Gaume and D. Z. Freedman, {\sl `Geometrical 
structure and Ultraviolet Finiteness in the Supersymmetric Sigma Model'}, 
Comm. Math. Phys. {\bf 80} (1981) 443.} 
\lref\myoldpaper{C. V. Johnson, {\sl `Exact Models of Extremal Dyonic 4D 
Black Hole Solutions of Heterotic String Theory'}, Phys. Rev. {\bf D50} (1994)
4032, hep-th/9403192.}
\lref\gojoe{J. Polchinski, {\sl `Dirichlet Branes and Ramond--Ramond Charges
 in String Theory'}, Phys. Rev. Lett. {\bf 75} (1995) hep-th/9510017.}
\lref\dnotes{J. Polchinski, S. Chaudhuri and C. V. Johnson, {\sl `Notes on 
D--Branes'}, hep-th/9602052.}
\lref\joetasi{J. Polchinski, `TASI Lectures on D-Branes', hep-th/9611050.}
\lref\hull{C. M.  Hull, {\sl `String--String Duality in Ten Dimensions'}, 
 Phys. Lett. {\bf B357} (1995) 545,  hep-th/9506194.}

\lref\dine{M. Dine, N. Seiberg and E. Witten, {\sl 
`Fayet--Iliopoulos Terms in String Theory'}, Nucl. Phys. {\bf B289}
(1987) 589.}
\lref\taylor{W. Taylor,
{\sl `D--Brane field theory on compact spaces'}, Phys. Lett. {\bf B394}
 (1997) 283, hep-th/9611042\semi O. J. Ganor, S. Ramgoolam, W. Taylor,
 {\sl `Branes, Fluxes and Duality in M(atrix)-Theory'},
 Nucl. Phys. {\bf B492} (1997) 191, hep-th/9611202.}

\lref\matrixeight{D. Lowe, {\sl `$E_8{\times}E_8$
 Instantons in Matrix Theory '}, hep-th/9709015\semi O. Aharony,
M. Berkooz, S. Kachru, and E. Silverstein, {\sl `Matrix Description of
$(1,0)$ Theories in Six Dimensions'}, hep-th/9709118.}
\lref\matrixheterotic{U. H. Danielsson, G. Ferretti, {\sl `The Heterotic
 Life of the D-particle'}, Int. J. Mod. Phys. {\bf A12} (1997) 4581,
hep-th/9610082\semi S. Kachru and E. Silverstein, {\sl `On Gauge
Bosons in the Matrix Model Approach to M Theory'}, Phys. Lett. {\bf
B396} (1997) 70, hep-th/9612162\semi L. Motl, {\sl `Quaternions and
M(atrix) theory in spaces with boundaries'}, hep-th/9612198\semi
N. Kim and S.-J. Rey, {\sl `M(atrix) Theory on an Orbifold and Twisted
Membrane'}, Nucl. Phys. {\bf B504} (1997) 189\semi D. Lowe, {\sl
`Bound States of Type I' D-particles and Enhanced Gauge Symmetry'},
Nucl. Phys. {\bf B501} (1997) 134, hep-th/9702006\semi T. Banks,
N. Seiberg, E. Silverstein, {\sl `Zero and One-dimensional Probes with
N=8 Supersymmetry'}, Phys.Lett. {\bf B401} (1997) 30,
hep-th/9703052\semi T. Banks and L. Motl, {\sl `Heterotic Strings from
Matrices'}, hep-th/9703218\semi D. Lowe, {\sl `Heterotic Matrix String
Theory'}, Phys. Lett. {\bf B403} (1997) 243, hep-th/9704041\semi
S-J. Rey, {\sl `Heterotic M(atrix) Strings and Their Interactions'},
hep-th/9704158 }
\lref\matrixheteroticii{P. Horava, {\sl `Matrix Theory and
Heterotic Strings on Tori'}, hep-th/9705055\semi D. Kabat and
S-J. Rey, {\sl `Wilson Lines and T-Duality in Heterotic M(atrix)
Theory'}, hep-th/9707099.} 
\lref\matrixheteroticiii{S. Govindarajan, {\sl `Heterotic
M(atrix) theory at generic points in Narain moduli space'},
hep-th/9707164.}
\lref\motl{L. Motl, {\sl `Proposals on Non--Perturbative Superstring
 Interactions'}, hep-th/9701025.}
\lref\banks{T. Banks and N. Seiberg, {\sl `Strings from Matrices'},
 hep-th/9702187.}
\lref\douglasooguri{M. R. Douglas, H. Ooguri, S. H. Shenker, 
{\sl `Issues in M(atrix) Theory Compactification'}, Phys.Lett. {\bf
 B402} (1997) 36, hep-th/9702203\semi M. R. Douglas, H. Ooguri {\sl
 `Why Matrix Theory is Hard'}, hep-th/9710178.}
\lref\fischler{W. Fischler, A. Rajaraman, 
{\sl `M(atrix) String Theory on K3'},
 hep-th/9704123.}
\lref\edmtheory{E. Witten, {\sl `Five-branes And 
$M$-Theory On An Orbifold'}, Nucl. Phys. {\bf B463} (1996) 383,
hep-th/9512219.}

\lref\sixteeneight{D. Morrison and C. Vafa,
{\sl `Compactifications of F--Theory on Calabi--Yau Threefolds -- I'},
Nucl.Phys. {\bf B473} (1996) 74, hep-th/9602114\semi D. Morrison and
C. Vafa, {\sl `Compactifications of F-Theory on Calabi--Yau Threefolds
-- II'}, Nucl. Phys. {\bf B476} (1996) 437, hep-th/9603161.}
\lref\ericper{P. Berglund  and E. Gimon, to appear.}
\lref\eric{E. Gimon, private communication. See also ref.\ericper.}
\lref\font{G. Aldazabal, A. Font, L.E. Ibanez, F. Quevedo, 
{\sl `Heterotic/Heterotic Duality in D=6,4'}, Phys.Lett. {\bf B380}
(1996) 33, hep-th/9602097.}

\centerline{\bf Contents}

\bigskip

\noindent {1.} {\fam \bffam \tenbf Motivation and Background} \leaderfill{3}
 \par
\noindent \quad{1.1.} {\fam \slfam \tensl Compactifying M--Theory} 
\leaderfill{3} \par 
\noindent \quad{1.2.} {\fam \slfam \tensl The DMW Point:
 Heterotic/Heterotic Duality} \leaderfill{4} \par 
\noindent \quad{1.3.} {\fam \slfam \tensl Relation to the GP Model} 
\leaderfill{4} \par 
\noindent \quad{1.4.} {\fam \slfam 
\tensl Searching for a Matrix Theory Representation} \leaderfill{6} \par 
\noindent \quad{1.5.} {Outline} \leaderfill{7} \par 
\noindent {2.} {\fam \bffam \tenbf Some Orientifold Models} \leaderfill{8}
 \par
\noindent \quad{2.1.} {\fam \slfam \tensl The GP Orientifold Model:
 Type  IB on $K3$} \leaderfill{8} \par 
\noindent \quad{2.2.} {\fam \slfam \tensl Another Orientifold Model:
 Type IA on $K3$} \leaderfill{10} \par 
\noindent \quad{2.3.} {\fam \slfam \tensl M--Theory on $K3{\times }{\cal I}$}
 \leaderfill{12} \par 

\noindent \quad\quad\quad{2.3.1.} {\fam \slfam \tensl Digression on D8--Branes 
and O8--Planes} \leaderfill{12} \par 
\noindent \quad\quad\quad{2.3.2.} {\fam \slfam \tensl Inclusion of $K3$, 
D4--Branes and O4--Planes} \leaderfill{13} \par 

\noindent \quad{2.4.} {\fam \slfam \tensl 
The $E_8{\times }E_8$ and $Spin(32)/{\hbox {{\fam 0\tenrm Z}\kern -.4em\hbox
 {\fam 0\tenrm Z}}}_2$ 
Heterotic Strings on $K3$} \leaderfill{15} \par 
\noindent \quad{2.5.} {\fam \slfam \tensl Geometrical Picture of the Dualities}
 \leaderfill{16} \par 
\noindent \quad{2.6.} {\fam \slfam \tensl Origin of 
Heterotic/Heterotic Duality} \leaderfill{17} \par 
\noindent \quad{2.7.} {\fam \slfam \tensl Heterotic/Heterotic Duality 
as Eleven Dimensional Electromagnetic Duality}
 \leaderfill{21} \par 
\noindent {3.} {\fam \bffam \tenbf Probing with D--Branes} 
\leaderfill{24} \par 
\noindent \quad{3.1.} {\fam \slfam \tensl Probing with D1--Branes: The Model}
 \leaderfill{25} \par 
\noindent \quad\quad\quad{3.1.1.} {\fam \slfam \tensl Supersymmetry}
 \leaderfill{25} \par
\noindent \quad\quad\quad{3.1.2.} {\fam \slfam \tensl The 1--1 Strings}
 \leaderfill{26} \par
\noindent \quad\quad\quad{3.1.3.} {\fam \slfam \tensl The 1--9 Strings}
 \leaderfill{28} \par
\noindent \quad\quad\quad{3.1.4.} {\fam \slfam \tensl The 1--5 Strings} 
\leaderfill{29} \par
\noindent \quad\quad\quad{3.1.5.} 
{\fam \slfam \tensl The 9--9, 5--5 and 5--9 Strings} 
\leaderfill{29} \par
\noindent \quad\quad\quad{3.1.6.} 
{\fam \slfam \tensl Two Puzzles, and Their Solution} 
\leaderfill{30} \par
\noindent \quad\quad\quad{3.1.7.} {\fam \slfam \tensl The Closed Strings}
 \leaderfill{33} \par
\noindent \quad{3.2.} {\fam \slfam \tensl Probing with D1--Branes:
 The Geometry} \leaderfill{35} \par 
\noindent \quad\quad\quad{3.2.1.} {\fam \slfam \tensl Instantons, ALE 
Spaces, HyperK\"ahler Quotients and ADHM Data} \leaderfill{35}
 \par
\noindent \quad\quad\quad{3.2.2.} {\fam \slfam \tensl The Moduli Space 
of GP Models and Probe Models} \leaderfill{37} \par
\noindent \quad\quad\quad{3.2.3.} {\fam \slfam \tensl The Special Point}
 \leaderfill{40} \par

\noindent \quad{3.3.}
 {\fam \slfam \tensl Probing with D0--branes and D5--branes} 
\leaderfill{42} \par 

\noindent \quad\quad\quad{3.3.1.} {\fam \slfam \tensl D0--Brane Probes} 
\leaderfill{42} \par

\noindent \quad\quad\quad{3.3.2.} {\fam \slfam \tensl D5--Brane Probes} 
\leaderfill{42} \par

\noindent {4.} {\fam \bffam \tenbf Some (Partial)  Matrix Theory Proposals} 
\leaderfill{43} \par 
\noindent \quad{4.1.} {\fam \slfam \tensl Matrix Theory of M--theory
 on $X{\times }{\cal I}$} \leaderfill{44} \par 
\noindent \quad{4.2.} {\fam \slfam \tensl Matrix Theory of the 
$E_8{\times}E_8$ Heterotic String on $X{\times}{\cal I}$}
\leaderfill{44} \par
\noindent \quad{4.1.} {\fam \slfam \tensl Matrix Theory of M--theory
 on $K3{\times }{\cal I}$ and its Heterotic Limits} \leaderfill{45} \par 
\noindent {5.} {\fam \bffam \tenbf Beyond the Special Point} 
\leaderfill{47} \par 
\noindent {6.} {\fam \bffam \tenbf Discussion and Conclusion} 
\leaderfill{47} \par 
\noindent \quad{6.1.} {\fam \slfam \tensl Summary} \leaderfill{47} \par 
\noindent \quad{6.2.} {\fam \slfam \tensl New Directions} \leaderfill{47} \par 
\noindent {\fam \bffam \tenbf Acknowledgements} \leaderfill{50} \par
\noindent {\fam \bffam \tenbf References} \leaderfill{51} \par

\vfill\eject

\newsec{\bf Motivation and Background}
\subsec{\sl Compactifying M--Theory}
Consider the case of M--theory compactified on the surface
$K3{\times}{\cal I}$, where $\cal I$ is a line interval of length
$L$. Consistent compactification\edmtheory\ requires the introduction of 24
M5--branes, transverse to the $K3{\times}{\cal I}$, in order to cancel
the resulting chiral anomaly in the six dimensional transverse space
$\IR^6$.  The anomaly owes its presence to essentially the curvature
of~$K3$, which has Pontryagin number 24 in the appropriate units. Put
another way, $K3$'s curvature produces 24 units of magnetic $A^{(3)}$
charge in the background, which may be canceled by the 24 units of
magnetic charge of the M5--branes\mfive, M--theory's basic dynamical
magnetic $A^{(3)}$ charge carriers at low energy\foot{Here, $A^{(3)}$
denotes the three--form potential of eleven--dimensional supergravity,
the low energy limit of M--theory.}.

As is now standard lore, in the Ho\u{r}ava--Witten limit ($L{\to}0$),
this compactification of M--theory yields a discrete {\sl family} of
compactifications of the $E_8{\times}E_8$ heterotic string
on~$K3$. The resulting (ten dimensional) heterotic string coupling is
related to the interval size as $\lambda{\sim}L^{3/2}$. 

There are two distinct categories of heterotic vacua obtained in this
way.  The first is ``ordinary'' heterotic string vacua, where all of
the massless spectrum can be accounted for in string perturbation
theory. Choices made for the distribution of M5--branes translate
directly into the choice of embedding $n_1$ of the instantons in one
$E_8$ and the other $n_2{=}24{-}n_1$ into the second $E_8$. This
corresponds, in M--theory, to the partitioning of the transverse
M5--branes such that there are $n_1$ at one endpoint of the interval
$\cal I$, and $n_2$ at the other, before taking the heterotic limit.

There are also heterotic string vacua with $n_T$ massless states in
their spectrum which transform as tensors. The chiral anomaly can
allow them as long as $n_1{+}n_2{+}n_T{=}24$. These are not to be
understood as completely perturbative heterotic vacua, but ones which
contain contributions from non--perturbative effects which make their
presence felt all the way down to the weak coupling limit.  In
M--theory ($L{\neq}0$), this seemingly mysterious statement is
realized explicitly in terms of making the choice $(n_1,n_2)$ for the
numbers of M5--branes at the respective ends of the interval, while
soaking up the rest of $K3$'s curvature--induced magnetic $A^{(3)}$
charge with $n_T$ M5--branes in the interior. The tensor massless
degrees of freedom living one on each M5--brane world volume, are now
free to contribute to the overall spectrum.  In taking the stringy
limit, that arrangement yields this distinct second class of heterotic
string vacua.

In both cases, the heterotic string itself arises from wrapping the
M2--brane\mtwo\ on the interval~$\cal I$, the boundary of the brane giving a
string in the transverse space. M--theory details completely
independent of $L$ ({\it i.e.,} residing on the ends of the interval
where the boundary of the M2--brane lives) like the instantons, become
identifiable as perturbative string physics in the $L{\to}0$
limit. Meanwhile, M--theory physics of the bulk (which is also the
bulk of the M2--brane) is not perturbatively described in the limit.

\subsec{\sl The DMW Point: Heterotic/Heterotic Duality}

Turning back to the $n_T{=}0$ cases, it is known\duff\ that there is
one special case where there exists a different heterotic limit.  One
first performs an eleven dimensional electromagnetic duality operation
with respect to $A^{(3)}$. This necessarily involves the exchange of
the M2--brane with an M5--brane oriented with four of its directions
on $K3$, and one in the $\IR^6$. (It therefore contributes nothing
to the anomaly mentioned earlier.) Sending the volume of the $K3$ to
zero results in a string in $\IR^6$ which is the heterotic string.

There is actually only one way to do this geometrically\duff, and this
is for choice $(12,12)$ distribution of the transverse
M5--branes. Only then can the $K3$ be seen to support no residual
magnetic charge in bulk and therefore be consistently shrunk away to
zero with an M5--brane wrapped on it\foot{Really, only a partial
argument was presented in ref.\duff; It was not clear there how
the~$K3$ in the resulting heterotic compactification reappears. This
M--theory argument will be completed in this paper.}.

That is the eleven dimensional geometrical origin of
heterotic/heterotic duality in six dimensions, defining the special
Duff--Minasian--Witten point\duff\ in the moduli space of
$E_8{\times}E_8$ heterotic string $K3$ compactifications.  This is the
(long anticipated\refs{\duffetal}) ${\cal N}{=}1$ string/string
duality in six dimensions which is a natural ancestor
of ${\cal N}{=}2$ string duality in four dimensions.

\subsec{\sl Relation to the GP Model}

Fortuitously, there exists yet another duality (discussed by Berkooz,
Leigh, Polchinski, Schwarz, Seiberg and Witten\berkoozi) which relates
the special (DMW) point to a type IIB
orientifold\dbranes\orientifolds\ model representing the type IB
string compactified on $K3$, called the Gimon--Polchinski
model\ericjoe\foot{A special point in the moduli space of the
Gimon--Polchinski (GP) models was also recognized as a consistent
spectrum by Bianchi and Sagnotti in ref.\sagnotti. This is the point
with largest possible gauge group $U(16){\times}U(16)$. In the
interests of historical accuracy, we might call this point the BS--GP
point, reserving the name GP for the more general class of models
found in ref.\ericjoe.}.

There are a number of important points to be made here about the
establishment of this duality. More correctly, the heterotic/heterotic
DMW point is dual to precisely that special point in the moduli space
of GP models where there might be some hope to make strong/weak
coupling duality contact with a perturbative heterotic string
compactified on~$K3$. (This point, which we'll call the ``special GP
model'' has no regions of spacetime where the dilaton has a gradient,
thereby promising a heterotic conformal field theory dual
description. Such orbifold conformal field theories were presented in
refs.\refs{\font,\berkoozi}.)  One might have anticipated such a
duality as a descendent of the ten dimensional strong/weak coupling
duality between the $SO(32)$ type~IB string and the $SO(32)$ heterotic
string. However, there is no straightforward geometrical reason why
such a heterotic dual to the GP model should have anything to do with
heterotic/heterotic duality: The ten dimensionally motivated
expectation should only generically predict duality to some $K3$
compactification of the $SO(32)$ heterotic string.

Part of the work  of ref.\berkoozi\ in showing that the special GP model was
indeed dual to the DMW model involved proving that
a much more complicated example of a phenomenon familiar in toroidal
heterotic compactifications occurs here: There is a T--duality
between the $(12,12)$ $K3$ compactified $E_8{\times}E_8$ heterotic
string and a $K3$ compactification of the $SO(32)$ heterotic
string. In order to make this correspondence, it has to be realized
that in the latter case, special choices are made for the instantons
which refer to their structure as $Spin(32)/\IZ_2$ instantons and not
merely $SO(32)$ ones.

Once this T--duality connection is made (a perturbative demonstration
which boils down to making careful choices of $Spin(32)/\IZ_2$ vector
bundles and choosing an appropriate Wilson line), the ten dimensional
expectation goes through and makes a connection between the special GP
model and the DMW point.

Indeed, there are some amusing facts which bolster the
connection\berkoozi: For example, it is easy to construct the string
solitons on the type~IB side which eventually become dual heterotic
strings on the heterotic side. They are simply the two distinct types
of D--string probe one can make in the six uncompactified
dimensions. One is simply the ordinary D1--brane in $\IR^6$, while the
other is made by wrapping a D5--brane on the $K3$ to give a D--string
in $\IR^6$. Note that in the GP model, these two D--string solitons
are T--dual to each other under inversion of the volume of the
(orbifold) $K3$. Under the strong/weak coupling duality, one of these
strings becomes light and becomes the heterotic string, while the
other becomes an ordinary heterotic stringy soliton. The Type~IB
model's $K3$ inversion T--duality symmetry, which exchanges the two
D--strings, can now be seen to be ``pulled back'' under the
type~IB/heterotic (on $K3$) map to the DMW heterotic/heterotic
duality, explaining the presence of a dual heterotic string.

In the first half of this paper, all of these orientifold facts will
be explicitly linked to the facts presented in the previous subsection
(1.2) by direct connection between the orientifold models and
M--theory on one side, and the heterotic models and M--theory on the
other side. In the process, a number of puzzling M--theory statements
in relation to heterotic/heterotic duality will be clarified and
extended.

\subsec{\sl Searching for a Matrix Theory Representation}

By this point in the discussion, the informed (but patient) reader may
wonder why we have gone to all of this trouble to recall the above
interesting collection of facts.  The goal of the present paper is to
try to begin to understand how to construct a matrix theory
representation of the $K3{\times}{\cal I}$ compactified M--theory. It
is not clear just how to do this directly from the original Matrix
theory prescription\bfss, which gave rise to a representation of
uncompactified M--theory. The technology of compactification on a
surface $X$ involves\refs{\bfss,\taylor}\ studying the large $N$ limit
of supersymmetric theories whose target space is a ``dual'' space
$\widetilde{X}$. In the case of toroidal compactifications on
$X{=}T^n$ where $n{<}4$, the prescription is completely given in terms
of $n{+}1$ dimensional $SU(N)$ Yang--Mills theory, and the ``dual''
space which will be its target space is simply the torus
$\widetilde{T}^n$ made by inverting the radii of all of the $n$
constituent circles.

Beyond that class of compactifications, the situation is less clear,
principally due to the fact that the ultraviolet behaviour of
Yang--Mills theory in higher dimensions is ill--defined. (Also, beyond
the case of tori, it is not clear what the ``dual'' surface should
be.)  In the case of~$T^4$, however, it was shown that certain facts
about field theory conspire to allow a sensible conjecture\rozali\
about the existence of a sensible Matrix theory definition in terms of
the large $N$ limit of a certain six dimensional superconformal fixed
point theory. The existence of the appropriate field theory was
confirmed in ref.\berkoozii, and the Matrix theory consequences,
obtained by placing it on $\widetilde{T}^5$, were studied further, by
testing that known type~II string $U$--duality symmetries are
predicted\berkoozii.

Even in that case, there is enough supersymmetry and geometrical
symmetry of the compactification space (and its dual) to sensibly
motivate a complete matrix description within the matrix theory
framework.  For the cases we would like to consider here, there is
less supersymmetry and geometrical symmetry of the target space to
help in finding a matrix theory representation.

One way to proceed might be to note that the compactification space
which we wish to consider is geometrically realizable (at special
points) by orbifolds of tori. Perhaps progress may be made by
considering orbifolding the matrix theory definition of M--theory on
$T^5$\refs{\rozali,\berkoozii}. The problem with this is of course
that we have only a partial understanding of what it means to directly
orbifold in either M--theory or matrix theory, and so this is not an
avenue of approach without potential wrong turns.

Our starting point is to go back to the spirit of the original matrix
theory representation of M--theory and find a well--understood
D0--brane background in string theory to endow with an eleven
dimensional interpretation, building a matrix theory from the
resulting world--volume theory. This approach is not without its
pitfalls either. The models we may construct in such a case will of
course not have nearly enough supersymmetry to allow the powerful
statements about the nature and existence of certain D0--brane bound
states at large $N$ ---of the type made in the original matrix theory
proposal\bfss--- to be made.  Therefore we only expect a partial definition
of the required matrix theory to arise from considering complicated
D0--brane backgrounds.

We might hope, however, that we can deduce certain properties of the
matrix theory we need by using as many facts as we can, from as many
sources as possible. This is the place where the heterotic/heterotic
discussion at the beginning comes in. At that special point in the
moduli space of M--theory backgrounds, we have a wealth of facts which
may help pin down the properties of the representation, (which we might
hope is some large $N$ limit of a special type of quantum  theory
compactified on a ``dual'' space to the $K3{\times}{\cal I}$, where
``dual'' needs to be defined.)

The very existence of the special properties of the DMW point might
have been motivation enough to try considering its consequences for a
matrix theory proposal\foot{Indeed, this was the question which led
the author to embark upon this project.}. As luck would have it though,
we are blessed with a complete D--brane model ---more precisely, an
orientifold model--- which serves as a representation of the special
heterotic/heterotic point.

We will therefore pursue our quest to find the required matrix model
representation by constructing a variant of the special point of the
GP model in the {\sl type~IA} theory. The matrix theory representation
of the compactification of M--theory on $K3{\times}{\cal I}$ will be
thus partially motivated as the large $N$ limit of a certain type IA
orientifold model probed by a conglomerate of $N$ D0--branes, and by
extension, also the large $N$ limit of D1-- and D5--brane probes of
type~IB orientifold models.

By the end of the discussion, we will have reason to conjecture that
more generally, the complete matrix theory representation of M--theory
on $K3{\times}{\cal I}$ is given by the large $N$ limit of a
compactification ---on a suitable ``dual'' surface--- of the ``little
$E_8{\times}E_8$ heterotic string'' ${\cal N}{=}1$ six dimensional
quantum theory.

\subsec{Outline}

In section~2 we begin by reminding the reader of some of the details
of the GP type~IIB orientifold model. We then introduce a new (but
closely related to the GP model) orientifold model in the type~IIA
theory, and translate the properties of the GP model into this new
model. We then study the strong coupling limit of this type~IA model,
which is an M--theory $K3{\times}{\cal I}$ compactification with
M--branes. From there we take a weak coupling limit which yields
heterotic string theory.

As promised, along the way we will recall, derive, rederive and
strengthen our knowledge of properties of the heterotic/heterotic
duality, in terms of its image dualities in the various dual contexts
of type~IB, type~IA and M--theory. (The interplay between the latter
two is the newly presented context.)  By the end of the section, we
will have thus firmly recast our knowledge of heterotic/heterotic duality
in type~IA/M--theory terms, in preparation for section~3.

In section~3, we take our type~I models and study them as backgrounds
for D--brane propagation. We consider therefore a model of quantum
mechanics, for the case of probing local parts of the $K3{\times}{\cal
I}$ geometry with D0--branes, a 1+1 dimensional theory when $L_{\rm
M}$ is small, resulting from D1--brane probes, and later a certain
$5+1$ dimensional quantum theory compactified on a dual surface to
$K3{\times}{\cal I}$ for the case of the full compactification.

In section~4 we submit these models as partial definitions of the
Matrix theory representation of M--theory on $K3{\times}{\cal I}$, at
the DMW point, in the infinite momentum frame. As these models inherit
the duality properties of the type~IA(B) and heterotic models which
they were derived from, they are extremely well suited to be used in
such a definition. However, due to the lack of enough supersymmetry,
there is not a great deal of control over the properties of the short
distance information (encoded in the bound state dynamics) which is
needed to supplement the model to make it a complete definition.

We discuss such shortcomings of the definition and
nevertheless present a proposal for the end result of the inclusion of the
missing details in the completion of the models of sections~3 and~4
into a satisfactory definition.

There are good reasons to believe, based on the properties of the
orientifold models, that the full 5+1 dimensional theory is the large
$N$ limit of the (0,1)$_N$ supersymmetric heterotic theory with
$E_8{\times}E_8$ global symmetry compactified on a dual surface to
$K3{\times}{\cal I}$.  At the DMW point it is also equivalent to a
compactification of the (0,1)$_N$ supersymmetric theory with
$Spin(32)/\IZ_2$ global symmetry.  (It is the infinite momentum frame
of the M--theory compactification which corresponds to a large $N$
limit of such a configuration.)

Following that identification, it is natural to conjecture (by
section~5) that the $E_8{\times}E_8$ theory can be used to construct
the matrix representation of M--theory on $K3{\times}{\cal I}$ even
away from the DMW point.

\newsec{\bf Some Orientifold Models}

\subsec{The GP Orientifold  Model: Type~IB on $K3$}

Let us begin by recalling some of the details of the Gimon--Polchinski
model\ericjoe.  It is a compactification of the type~IB string on a $\IZ_2$
orbifold of $T^4$, realized as the prototype $K3$ orientifold of the
type~IIB string\foot{See refs.\refs{\ericmeI,\atish} for a wider discussion of
$K3$ orientifolds of type~IIB.}, where spacetime orbifold symmetries
are combined with worldsheet (orientifold) ones. The orientifold
model, defined by the group 
\eqn\orientifoldG{G_\Omega:=\{1,R_{6789},\Omega,\Omega
R_{6789}\},} contains the orbifold symmetry
\eqn\reflection{{R_{6789}:\, \{x^6,x^7,x^8,x^9\}\to\{-x^6,-x^7,-x^8,-x^9\},}}
and the orientifold symmetry ${\Omega:\, z\leftrightarrow {\bar z}}$
on the world--sheet which is parameterized by complex coordinate $z$
at string tree level.

The presence of $\Omega$ in the group may be thought of as
representing a single orientifold plane (``O9--plane'') which fills
the whole of spacetime. This plane has $-32$ units of D9--brane
charge. To cancel this charge (as required by ``Gauss' Law'' for the
Ramond--Ramond potential $A^{(10)}$), 32 D9--branes are introduced.

The presence of $\Omega R_{6789}$ represents the introduction of $16$
O5--planes, located one at each of the orbifold fixed points of the
singular $T^4/\IZ_2$ geometry.  Their $-2$ units each of D5--brane
charge require 32 D5--branes to be introduced, for consistent
$A^{(6)}$ field equations.

In the top part of Table~1 is illustrated the orientations of all of
the constituents of the model. The $K3{=}T^4/\IZ_2$ is located in the
$\{x^6,x^7,x^8,x^9\}$ directions. The various D5--branes are free to
move around in the $K3$, but only in groups of four\ericjoe, as there
must be a mirror partner of each brane under $R_{6789}$, and there is
an additional pairing for invariance under $\Omega$. We may also
introduce Wilson lines to break some of the gauge symmetry of the
D5--branes and D9--branes.

\midinsert{
\bigskip
\vbox{
$$\vbox{\offinterlineskip
\hrule height 1.1pt
\halign{&\vrule width 1.1pt#
&\strut\quad#\hfil\quad&
\vrule#
&\strut\quad#\hfil\quad&
\vrule width 1.1pt#
&\strut\quad#\hfil\quad&
\vrule#
&\strut\quad#\hfil\quad&
\vrule#
&\strut\quad#\hfil\quad&
\vrule#
&\strut\quad#\hfil\quad&
\vrule#
&\strut\quad#\hfil\quad&
\vrule#
&\strut\quad#\hfil\quad&
\vrule\hskip 2pt
\vrule#
&\strut\quad#\hfil\quad&
\vrule#
&\strut\quad#\hfil\quad&
\vrule#
&\strut\quad#\hfil\quad&
\vrule#
&\strut\quad#\hfil\quad&
\vrule width 1.1pt#\cr
height3pt
&\omit&
&\omit&
&\omit&
&\omit&
&\omit&
&\omit&
&\omit&
&\omit&
&\omit&
&\omit&
&\omit&
&\omit&
\cr
&\hfil type&
&\hfil \#&
&\hfil $x^0$&
&\hfil $x^1$&
&\hfil $x^2$&
&\hfil $x^3$&
&\hfil $x^4$&
&\hfil $x^5$&
&\hfil $x^6$&
&\hfil $x^7$&
&\hfil $x^8$&
&\hfil $x^9$&
\cr
height3pt
&\omit&
&\omit&
&\omit&
&\omit&
&\omit&
&\omit&
&\omit&
&\omit&
&\omit&
&\omit&
&\omit&
&\omit&
\cr
\noalign{\hrule height 1.1pt}
height3pt
&\omit&
&\omit&
&\omit&
&\omit&
&\omit&
&\omit&
&\omit&
&\omit&
&\omit&
&\omit&
&\omit&
&\omit&
\cr
&\hfil D9&
&\hfil $32$&
&\hfil --- &
&\hfil --- &
&\hfil --- &
&\hfil --- &
&\hfil --- &
&\hfil --- &
&\hfil --- &
&\hfil --- &
&\hfil --- &
&\hfil --- &
\cr
height3pt
&\omit&
&\omit&
&\omit&
&\omit&
&\omit&
&\omit&
&\omit&
&\omit&
&\omit&
&\omit&
&\omit&
&\omit&
\cr
\noalign{\hrule}
height3pt
&\omit&
&\omit&
&\omit&
&\omit&
&\omit&
&\omit&
&\omit&
&\omit&
&\omit&
&\omit&
&\omit&
&\omit&
\cr
&\hfil O9&
&\hfil $1$&
&\hfil --- &
&\hfil --- &
&\hfil --- &
&\hfil --- &
&\hfil --- &
&\hfil --- &
&\hfil --- &
&\hfil --- &
&\hfil --- &
&\hfil --- &
\cr
height3pt
&\omit&
&\omit&
&\omit&
&\omit&
&\omit&
&\omit&
&\omit&
&\omit&
&\omit&
&\omit&
&\omit&
&\omit&
\cr
\noalign{\hrule}
height3pt
&\omit&
&\omit&
&\omit&
&\omit&
&\omit&
&\omit&
&\omit&
&\omit&
&\omit&
&\omit&
&\omit&
&\omit&
\cr
&\hfil D5&
&\hfil $32$&
&\hfil --- &
&\hfil --- &
&\hfil --- &
&\hfil --- &
&\hfil --- &
&\hfil --- &
&\hfil $\bullet$ &
&\hfil $\bullet$ &
&\hfil $\bullet$ &
&\hfil $\bullet$ &
\cr
height3pt
&\omit&
&\omit&
&\omit&
&\omit&
&\omit&
&\omit&
&\omit&
&\omit&
&\omit&
&\omit&
&\omit&
&\omit&
\cr
\noalign{\hrule}
height3pt
&\omit&
&\omit&
&\omit&
&\omit&
&\omit&
&\omit&
&\omit&
&\omit&
&\omit&
&\omit&
&\omit&
&\omit&
\cr
&\hfil O5&
&\hfil $16$&
&\hfil --- &
&\hfil --- &
&\hfil --- &
&\hfil --- &
&\hfil --- &
&\hfil --- &
&\hfil $\bullet$ &
&\hfil $\bullet$ &
&\hfil $\bullet$ &
&\hfil $\bullet$ &
\cr
height3pt
&\omit&
&\omit&
&\omit&
&\omit&
&\omit&
&\omit&
&\omit&
&\omit&
&\omit&
&\omit&
&\omit&
&\omit&
\cr
\noalign{\hrule}
height2pt
&\omit&
&\omit&
&\omit&
&\omit&
&\omit&
&\omit&
&\omit&
&\omit&
&\omit&
&\omit&
&\omit&
&\omit&
\cr
\noalign{\hrule}
height3pt
&\omit&
&\omit&
&\omit&
&\omit&
&\omit&
&\omit&
&\omit&
&\omit&
&\omit&
&\omit&
&\omit&
&\omit&
\cr
&\hfil D1&
&\hfil $N$&
&\hfil --- &
&\hfil $\bullet$ &
&\hfil $\bullet$ &
&\hfil $\bullet$ &
&\hfil $\bullet$ &
&\hfil --- &
&\hfil $\bullet$ &
&\hfil $\bullet$ &
&\hfil $\bullet$ &
&\hfil $\bullet$ &
\cr
height3pt
&\omit&
&\omit&
&\omit&
&\omit&
&\omit&
&\omit&
&\omit&
&\omit&
&\omit&
&\omit&
&\omit&
&\omit&
\cr
\noalign{\hrule}
height3pt
&\omit&
&\omit&
&\omit&
&\omit&
&\omit&
&\omit&
&\omit&
&\omit&
&\omit&
&\omit&
&\omit&
&\omit&
\cr
&\hfil D5$^\prime$&
&\hfil $N$&
&\hfil --- &
&\hfil $\bullet$ &
&\hfil $\bullet$ &
&\hfil $\bullet$ &
&\hfil $\bullet$ &
&\hfil --- &
&\hfil --- &
&\hfil --- &
&\hfil --- &
&\hfil --- &
\cr
height 3pt
&\omit&
&\omit&
&\omit&
&\omit&
&\omit&
&\omit&
&\omit&
&\omit&
&\omit&
&\omit&
&\omit&
&\omit&
\cr
}\hrule height 1.1pt
}
$$
 {\bf Table 1.} {\it (Top) Orientation and number of the consituent
branes of the Gimon--Polchinski model of the type~IB string
compactified on a $T^4/\IZ_2$ orbifold limit of $K3$ (located in the
6,7,8 and 9 directions). (Bottom) Orientation of branes used to probe
the model (see section~3).}  } }
\endinsert

The gauge symmetry comes from two sectors, the 9--9 sector and the
5--5 sector, and is maximal when there are no Wilson lines, and the
D5--branes are all coincident located on one of the 16 fixed
points. This gives gauge group\foot{As shown in ref.\berkoozi, the
$U(16)$'s are broken to $SU(16)$ by one--loop  effects. See later.}\
$U(16)_5{\times}U(16)_9$. (The subscripts denote a gauge group's
origin in the 9--9 or 5--5 sector.) 9--9 and 5--5 strings also supply
hypermultiplets in the $\bf{120}{+}\overline{\bf{120}}$ for each
factor, and 9--5 strings supply a $({\bf16},{\bf16})$. The closed
string sector supplies 20 neutral hypermultiplets representing the
(stringy) geometric deformations of $K3$\foot{This special point is
the BSGP point referred to earlier\sagnotti\ericjoe}.

As the number of D5--branes which may move around the $K3$ is limited
to be a multiple of four, there are a number of distinct models which
may be constructed in the GP class which are not connected\ericjoe
(see a reminder in section~3.2.2). One such model is the special point
we are interested in: It has a pair of D5--branes at each O5--plane,
thereby canceling the $A^{(6)}$ {\sl locally}. This unique arrangement
is important for the dualities which we want to consider because such
a local cancellation in the Ramond--Ramond sector implies a local
cancellation in the Neveu-Schwarz--Neveu-Schwarz sector, thus ensuring
that the dilaton is constant everywhere. Consequently, any strong/weak
dual heterotic model is likely to be realizable as a conformal field
theory, as demonstrated explicitly in ref.\berkoozi.

For this configuration, there is no gauge group, and there are simply
224 hypermultiplets from the open string sector plus the 20 neutral
ones from the closed string sector. (More details on this are to be
found in section~3.2.3.)

As pointed out in ref.\ericjoe, the four D5--brane dynamical unit
constitutes one complete instanton. Therefore there are 8 such
instantons in the D5--brane sector, the other 16 coming from the 16
fixed points to make a total of 24, as required. It is crucial to
note\berkoozi\ that these are $Spin(32)/\IZ_2$ instantons ``without
vector structure'', which means that the gauge bundle over the fixed
points is chosen in such a way that $U(1)$ instantons can be embedded
into $Spin(32)/\IZ_2$ and yield the correct Dirac quantization for the
adjoint and (say) positive chirality spinor representations, but not
for the vector and negative chirality spinor representations. It is
$Spin(32)/\IZ_2$ heterotic string theory compactified on a $K3$ with
such instantons which is T--dual to a $K3$ compactification of the
$E_8{\times}E_8$ heterotic string, as will be useful for us later.  We
will also see how these instantons arise naturally in the D--brane
probe models in section~3.

\subsec{Another Orientifold Model: Type~IA on $K3$.}

Table 2 shows an arrangement of branes and orientifolds in the type~IA
theory. 

The model has orientifold group
\eqn\orientifoldG{G_\Omega=\{1,R_{6789},\Omega R_5,\Omega R_{56789}\}.}
The compactification is geometrically essentially
$K3(\IZ_2){\times}S^1/\IZ_2$ where the $S^1$ is a compact $x^5$ of
radius $L^5_{\rm IA}/2\pi$. The $\IZ_2$ is generated by $R_5:\,
x^5{\to}-x^5$. The presence of element $\Omega R_5$ indicates the
presence of two orientifold O8--planes, at the two $\IZ_2$ fixed
points of the orbifolded circle, forming an interval~${\cal I}$. They
have charge $-16$ each in D8--brane units and therefore there are 32
D8--branes for the model to be consistent.

Meanwhile $\Omega R_{56789}$ requires $32$ O4--planes, split as
$16{\times}2$. The sixteen are associated with the fixed points of the
$T^4/\IZ_2$ $K3$ orbifold, and there are two copies of this
arrangement, one for each of the $x^5$ fixed points, the ends of the
interval~${\cal I}$.

Each O4--plane has $-1$ units of D4--brane charge and so there are 32
D4--branes in the model for consistency. They are localized in the
five dimensional $K3{\times}{\cal I}$ space.

\midinsert{
\bigskip
\vbox{
$$\vbox{\offinterlineskip
\hrule height 1.1pt
\halign{&\vrule width 1.1pt#
&\strut\quad#\hfil\quad&
\vrule#
&\strut\quad#\hfil\quad&
\vrule width 1.1pt#
&\strut\quad#\hfil\quad&
\vrule#
&\strut\quad#\hfil\quad&
\vrule#
&\strut\quad#\hfil\quad&
\vrule#
&\strut\quad#\hfil\quad&
\vrule#
&\strut\quad#\hfil\quad&
\vrule#
&\strut\quad#\hfil\quad&
\vrule
\hskip 2pt
\vrule#
&\strut\quad#\hfil\quad&
\vrule#
&\strut\quad#\hfil\quad&
\vrule#
&\strut\quad#\hfil\quad&
\vrule#
&\strut\quad#\hfil\quad&
\vrule width 1.1pt#\cr
height3pt
&\omit&
&\omit&
&\omit&
&\omit&
&\omit&
&\omit&
&\omit&
&\omit&
&\omit&
&\omit&
&\omit&
&\omit&
\cr
&\hfil type&
&\hfil \#&
&\hfil $x^0$&
&\hfil $x^1$&
&\hfil $x^2$&
&\hfil $x^3$&
&\hfil $x^4$&
&\hfil $x^5$&
&\hfil $x^6$&
&\hfil $x^7$&
&\hfil $x^8$&
&\hfil $x^9$&
\cr
height3pt
&\omit&
&\omit&
&\omit&
&\omit&
&\omit&
&\omit&
&\omit&
&\omit&
&\omit&
&\omit&
&\omit&
&\omit&
\cr
\noalign{\hrule height 1.1pt}
height3pt
&\omit&
&\omit&
&\omit&
&\omit&
&\omit&
&\omit&
&\omit&
&\omit&
&\omit&
&\omit&
&\omit&
&\omit&
\cr
&\hfil D8&
&\hfil $32$&
&\hfil --- &
&\hfil --- &
&\hfil --- &
&\hfil --- &
&\hfil --- &
&\hfil $\bullet$ &
&\hfil --- &
&\hfil --- &
&\hfil --- &
&\hfil --- &
\cr
height3pt
&\omit&
&\omit&
&\omit&
&\omit&
&\omit&
&\omit&
&\omit&
&\omit&
&\omit&
&\omit&
&\omit&
&\omit&
\cr
\noalign{\hrule}
height3pt
&\omit&
&\omit&
&\omit&
&\omit&
&\omit&
&\omit&
&\omit&
&\omit&
&\omit&
&\omit&
&\omit&
&\omit&
\cr
&\hfil O8&
&\hfil $2$&
&\hfil --- &
&\hfil --- &
&\hfil --- &
&\hfil --- &
&\hfil --- &
&\hfil $\bullet$ &
&\hfil --- &
&\hfil --- &
&\hfil --- &
&\hfil --- &
\cr
height3pt
&\omit&
&\omit&
&\omit&
&\omit&
&\omit&
&\omit&
&\omit&
&\omit&
&\omit&
&\omit&
&\omit&
&\omit&
\cr
\noalign{\hrule}
height3pt
&\omit&
&\omit&
&\omit&
&\omit&
&\omit&
&\omit&
&\omit&
&\omit&
&\omit&
&\omit&
&\omit&
&\omit&
\cr
&\hfil D4&
&\hfil $32$&
&\hfil --- &
&\hfil --- &
&\hfil --- &
&\hfil --- &
&\hfil --- &
&\hfil $\bullet$ &
&\hfil $\bullet$ &
&\hfil $\bullet$ &
&\hfil $\bullet$ &
&\hfil $\bullet$ &
\cr
height3pt
&\omit&
&\omit&
&\omit&
&\omit&
&\omit&
&\omit&
&\omit&
&\omit&
&\omit&
&\omit&
&\omit&
&\omit&
\cr
\noalign{\hrule}
height3pt
&\omit&
&\omit&
&\omit&
&\omit&
&\omit&
&\omit&
&\omit&
&\omit&
&\omit&
&\omit&
&\omit&
&\omit&
\cr
&\hfil O4&
&\hfil $32$&
&\hfil --- &
&\hfil --- &
&\hfil --- &
&\hfil --- &
&\hfil --- &
&\hfil $\bullet$ &
&\hfil $\bullet$ &
&\hfil $\bullet$ &
&\hfil $\bullet$ &
&\hfil $\bullet$ &
\cr
height3pt
&\omit&
&\omit&
&\omit&
&\omit&
&\omit&
&\omit&
&\omit&
&\omit&
&\omit&
&\omit&
&\omit&
&\omit&
\cr
\noalign{\hrule}
height2pt
&\omit&
&\omit&
&\omit&
&\omit&
&\omit&
&\omit&
&\omit&
&\omit&
&\omit&
&\omit&
&\omit&
&\omit&
\cr
\noalign{\hrule}
height3pt
&\omit&
&\omit&
&\omit&
&\omit&
&\omit&
&\omit&
&\omit&
&\omit&
&\omit&
&\omit&
&\omit&
&\omit&
\cr
&\hfil D0&
&\hfil $N$&
&\hfil --- &
&\hfil $\bullet$ &
&\hfil $\bullet$ &
&\hfil $\bullet$ &
&\hfil $\bullet$ &
&\hfil $\bullet$ &
&\hfil $\bullet$ &
&\hfil $\bullet$ &
&\hfil $\bullet$ &
&\hfil $\bullet$ &
\cr
height3pt
&\omit&
&\omit&
&\omit&
&\omit&
&\omit&
&\omit&
&\omit&
&\omit&
&\omit&
&\omit&
&\omit&
&\omit&
\cr
}\hrule height 1.1pt
}
$$
{\bf Table 2.} {\it (Top) Number and configuration of constituent
branes in an orientifold model of the type~IA string compactified on a
$T^4/\IZ_2$ orbifold limit of $K3$ (located in the 6, 7, 8 and 9
directions) times the line interval $S^1/\IZ_2$ (located in the 5
direction). (Bottom) Orientation of D0--brane used to probe the model
locally realizing a model of matrix quantum mechanics (see
section~3).} } }
\endinsert

It should be straightforward to see that this model is essentially
$T_5$--dual to the GP model compactified on an $x^5$ circle of radius
$L^5_{\rm IB}/2\pi$, where the relation between parameters of the two
models is\foot{Here and in what follows, we are not concerned with
factors of $\alpha^\prime$; nor are we careful about factors of
$2\pi$. These will often be omitted.}:
\eqn\paramsi{\eqalign{L^5_{\rm IB}&={1\over L^5_{\rm IA}}\cr
\lambda_{\rm IB}&={\lambda_{\rm IA}\over L^5_{\rm IA}}\cr
R_{\rm IB}&=R_{\rm IA},}} where $\lambda_{IA(B)}$ denotes the string coupling
and $R_{IA(B)}$ denotes the radius of the $K3$ manifold in the
type~IA(B) theory.

\subsec{\sl M--Theory on $K3{\times}{\cal I}$}

Just as in the case of the GP model, there is a particularly
interesting unique configuration where all of the R--R charges are
completely canceled locally in the model. The required arrangement is
the placement of 16 of the D8--branes on an O8--plane at one end of
the interval and the remaining 16 on the O8--plane at the other
end\horavawitten. Meanwhile the D4--branes are distributed one per
O4--plane, resulting in 16 at each of the copies of $K3$ at the end of
the interval.

$\underline{\hbox{\sl 2.3.1 Digression on  D8--branes  and O8--planes}}$

\medskip

In the absence of $K3$, the D4--branes and the O4--planes, this
arrangement results in gauge group $SO(16){\times}SO(16)$ for the
type~IA theory, resulting from the 16 D8--branes and one O8--plane at
each end of the interval.  As pointed out in
refs.\refs{\horavawitten,\joetasi}, this unique arrangement is
suggestive: Away from the world volumes of the branes, {\it i.e.,} in
the bulk of the $x^5$ interval, the model is locally the physics of
the type~IIA string. As we go to strong coupling in this model
therefore, it is to be expected that a new dimension, $x^{10}$, opens
up, and the model becomes an eleven dimensional theory. Each of the 16
D8--branes and O8--plane arrangements becomes a single object at
strong coupling, supplying the two ends of the Ho\u{r}ava--Witten
compactification of M--theory, becoming a sort of `nine--brane' after
unwrapping a leg in the $x^{10}$ direction. Crucially, soliton
states\dnotes\ which transform in the ${\bf(1,128)}+{\bf(128,1)}$
({\it i.e.,} the spinor of $SO(16){\times}SO(16)$) become light in
this limit, joining the $\bf{(1,120)}+\bf{(120,1)}$ in filling out
each $SO(16)$ into $E_8$.
 
In fact, one can go further than this and see that {\sl any other
arrangement} of D8--branes and O8--planes will not produce an eleven
dimensional theory. This issue is intimately related to the matter of
when it is consistent to have D8--branes in type~IIA theory\foot{I
have benefited from conversations with R. Myers concerning the
O8--plane matters discussed in the rest of this sub--subsection.}.

An isolated D8--brane in type~IIA string theory is not consistent. As
pointed out in refs.\refs{\joetasi\others}, the linear behaviour of
the dilaton with the transverse distance is such that the string
coupling diverges a {\sl finite} distance away from the brane. In
order to fix this state of affairs, it is necessary to introduce an
object which acts as a sink for the dilaton field, cutting off the
growth of the string coupling at some chosen reference finite value,
$\lambda^c_{\rm IA}$.  By supersymmetry, the object with the correct
charges to do this is the object with opposite R--R charge to the
D8--brane, the O8--plane.

Once we have made a reference choice for the value $\lambda^c_{\rm
IA}$ at which the type~IIA string coupling should be cut off, we have
fixed the distance that the O8--plane is from the D8--brane. Due to
the $-16$ units of charge of an O8--plane, overall consistency
requires there to be 16 D8--branes. On introduction of the O8--planes
and D8--branes, we are no longer in the type~IIA theory (globally) but
the type~IA theory. Between each D8--brane, the theory is a piece of
type~IIA string theory with a dilaton gradient. (At low energy, it is
simply the massive type~IIA supergravity of
Romans\refs{\romans,\gojoe}.). We have constructed the physics of the
neighbourhood of an endpoint of the type~IA interval ${\cal I}$.

So the answer to the question ``When is it consistent to have a
D8--brane in type~IIA theory?'' has the answer ``When it is in the
type~IA string theory''.

Now if we chose to increase the string coupling, and try to study what
happens at strong coupling, the situation is interesting. In order to
keep the theory (in the limit of infinite coupling) from developing
isolated regions of infinite string coupling again, we preserve our
chosen reference cutoff value $\lambda^c_{\rm IA}$ by simply moving the
orientifolds closer to the D8--branes, as the coupling increases. We
can do this indefinitely. However, there comes a point at which the
interval upon which all of this physics is taking place is forced by
this process to become so short (compared to $\sqrt{\alpha^\prime}$)
that we make stringy sense of it by simply T--dualizing along that
interval to go to the type~IB string where the distances in the dual
direction are large compared to $\sqrt{\alpha^\prime}$.

The resulting physics of the continued process of increasing the
string coupling further until it is infinitely strong, is addressed
succinctly by going to the dual theory to the type~IB string --- the
$SO(32)$ heterotic string.

This process of increasing the string coupling in the type~IA theory
will always result ---for any configuration of D8--branes and
O8--planes except one--- in the weakly coupled physics of the $SO(32)$
heterotic string in the limit, because the O8--planes and D8--branes
will always have to be moved closer to one another until the physics
is best addressed in terms of strong coupling type~IB physics.

The exception to this situation is the unique case where the dilaton
charges are canceled locally, as mentioned a few paragraphs
hence. In that case, there is no need to move O8--planes (reducing the
length of the interval) to cutoff regions of infinite coupling. The
strong coupling physics of that configuration is related to that of
the $E_8{\times}E_8$ heterotic string, as discussed above.

$\underline{\hbox{\sl 2.3.2. Inclusion of $K3$, D4--Branes and
O4--Planes}}$

\medskip 

Returning to the $K3$ compactification, we see that the discussion of
the last section has a counterpart here. In the compactified model, we
have also to include D4--branes to cancel the charge of the O4--plane
fixed points on $K3$.

In analogy with  similar processes as for D--branes, one may think of
the origin of the O4--planes as arising from wrapping the
O8--plane\foot{This way of thinking of O4--planes was suggested to me
by R. Myers} on the $K3$, whose curvature is localized at its fixed
points in this orbifold limit. $K3$'s non--zero value of $R\wedge R$
(supported at each fixed point) in the worldvolume action of the
O8--plane endows each fixed point with O4--plane charge.

This necessity to introduce D4--branes and O4--planes is entirely
analogous to the compactification of M--theory on $K3{\times}{\cal
I}$, or the $E_8{\times}E_8$ heterotic string on $K3$. The curvature
of $K3$ requires the introduction of 24 instantons for consistency. In
this type~IA situation, the instantons on $K3$ are played by
combinations of O4--planes and D4--branes. In the strong coupling
limit, where we make contact with M--theory on $K3{\times}{\cal I}$,
we expect them to become five--brane like objects (they reveal a
hidden leg wrapped around the new dimension $x^{10}$ which opens up),
destined to become $E_8$ instantons in the heterotic limit. This is
the analogue of the way that the combination of 16 D8--branes and 1
O8--plane became the M--theory ``nine--brane'' boundary carrying the
$E_8$ degrees of freedom. An isolated dynamical unit of four
D4--branes would become an ordinary M5--brane, while inside the $K3$,
a single O4--plane is related to half an M5--brane localized on a
fixed point of $K3$.

Recalling that this particular orientifold model is expected to become
the $(12,12)$ heterotic string compactification, we may deduce that it
is precisely 16 D4--branes plus 16 O4--planes which together become 12
$E_8$ instantons: 1 D4 and 1 O4 is equivalent to 3/4 an $E_8$
instanton. In particular, the O4--plane is equivalent to 1/2 of an
instanton, and the single D4--brane sitting in it in our special
configuration is 1/4 of an instanton. This is consistent with the
$T_5$--duality relation to the GP model and its relation to the
$Spin(32)/\IZ_2$ heterotic string compactification: There, four
D5--branes make up one dynamical unit which is dual to a small
heterotic instanton, a numerical relationship which should be
invariant under $T_5$--duality. Meanwhile, a single O5--plane is
equivalent to a heterotic instanton, which is consistent with our
type~IA assignment above, and the fact that $T_5$--duality turns two
O4--planes into one O5--plane.

The relationship between the parameters of the M--theory
compactification and those of type~IA may be easily computed to be
(see footnote~8):
\eqn\paramsii{\eqalign{
L^{10}_{\rm M}&=\lambda_{\rm IA}^{2\over3}\cr L^5_{\rm
M}&=\lambda_{\rm IA}^{-{1\over3}}L^5_{\rm IA}\cr R_{\rm
M}&=\lambda_{\rm IA}^{-{1\over3}}R_{\rm IA}, }} where $L^{10}_{\rm
M}/2\pi$, $L^5_{\rm M}/2\pi$ and $R_{\rm M}$ are the radii of $x^{10}$
direction, the $x^5$ direction and the~$K3$, measured in M--theory
length units. These relationships are simple consequences of
dimensional reduction of the eleven dimensional supergravity
Lagrangian on a circle to ten dimensions, and then performing a Weyl
rescaling to match the resulting Lagrangian to the string frame
Lagrangian of the massless sector of the relevant string theory\goed.

Using the relations \paramsi, we can compute the relationship between
the M--theory parameters and the type~IB parameters:
\eqn\paramsiii{\eqalign{
L^{10}_{\rm M}&={\lambda_{\rm IB}^{2\over3}\over(L^5_{\rm
IB})^{2\over3}}\cr L^5_{\rm M}&={1\over\lambda_{\rm
IB}^{{1\over3}}(L^5_{\rm IB})^{2\over3}}\cr R_{\rm M}&={R_{\rm
IB}(L^5_{\rm IB})^{1\over3}\over\lambda_{\rm IB}^{{1\over3}}}. }}

\subsec{The $E_8{\times}E_8$ and $Spin(32)/\IZ_2$ Heterotic 
Strings on $K3$}

At this stage we can contemplate moving back to six dimensions and
recover the $E_8{\times}E_8$ heterotic string compactified on $K3$.
This is performed by starting with our M--theory configuration and
shrinking the length of the interval by sending $L^5_{\rm
M}{\to}0$. (Shrinking $L^{10}_{\rm M}{\to}0$ would return us to the
six dimensional theory of the type~IA compactification.)

In the limit, we obtain the (12,12) heterotic compactification. This
is guaranteed by fact that the in type~IA orientifold model we chose
an arrangement of branes which is completely symmetric between the
ends of the interval. The tracing that we did of the instanton
assignments from their role as $Spin(32)/\IZ_2$ instantons in the
type~IB model, through the type~IA model, and on to the M--theory
configuration is therefore correct.

The relationship between the parameters of the heterotic model and
those of the M--theory compactification are:
\eqn\paramsiv{\eqalign{
L^{5}_{\rm M}&=\lambda_{\rm HA}^{2\over3}\cr L^{10}_{\rm
M}&=\lambda_{\rm HA}^{-{1\over3}}L^{10}_{\rm HA}\cr R_{\rm
M}&=\lambda_{\rm HA}^{-{1\over3}}R_{\rm HA}, }} where $L^{10}_{\rm
HA}/2\pi$, $L^5_{\rm HA}/2\pi$ and $R_{\rm HA}$ are the radii of
$x^{10}$ direction, the $x^5$ direction and the~$K3$, measured in
$E_8{\times}E_8$ heterotic string length units, while $\lambda_{\rm
HA}$ denotes that heterotic string's coupling\foot{For succinct
subscripts it is simpler ---and mindful of type~I/heterotic duality---
to use ``HA'' to denote the $E_8{\times}E_8$ heterotic string and
``HB'' to denote the $Spin(32)/\IZ_2$ heterotic string.}.

Finally we arrive at the heterotic model with the
assignment of instantons we expected from considerations\berkoozi\ of
$Spin(32)/\IZ_2$ heterotic/type~IB strong/weak coupling duality
combined with $T_{10}$--duality between the two $K3$ compactified
types of heterotic strings.

The $T_{10}$--duality gives the $Spin(32)/\IZ_2$ heterotic string with
parameters
\eqn\paramsv{\eqalign{L^{10}_{\rm HB}&={1\over L^{10}_{\rm HA}}\cr
\lambda_{\rm HB}&={\lambda_{\rm HA}\over L^{10}_{\rm HA}}\cr
R_{\rm HB}&=R_{\rm HA},}} where the notation should now be clear.

Combining this with \paramsiv\ gives a relationship between the
$Spin(32)/\IZ_2$ heterotic parameters and those of M--theory:
\eqn\paramsvi{\eqalign{
L^{5}_{\rm M}&=
{\lambda_{\rm HB}^{2\over3}\over(L^{10}_{\rm HB})^{2\over3}}\cr
L^{10}_{\rm M}&=
{1\over\lambda_{\rm HB}^{{1\over3}}(L^{10}_{\rm HB})^{2\over3}}\cr
R_{\rm M}&={R_{\rm HB}(R^{10}_{\rm IB})^{1\over3}\over
\lambda_{\rm HB}^{{1\over3}}}. }}

Comparing this to \paramsiii, if we did not know it already we could
deduce\horavawitten\ a strong/weak coupling duality between the
type~IB and the $Spin(32)/\IZ_2$ heterotic string with the precise
relationship\refs{\goed,\edjoe,\hull}:
\eqn\paramsvii{\eqalign{
\lambda_{\rm IB}&=\lambda^{-1}_{\rm HB}\cr
L^5_{\rm HB}&=L^{10}_{\rm HB}\lambda_{\rm HB}^{-{1\over2}}, }}
supplemented by the following relationship between the radii of the
$K3$'s upon which they are compactified:
\eqn\radii{R_{\rm HB}=R_{\rm IB}\lambda_{\rm IB}^{-{1\over2}}.}

For completeness, we also deduce a relationship between the type~IA
and $E_8{\times}E_8$ parameters:
\eqn\paramsix{\eqalign{
\lambda_{\rm IA}&=\lambda_{\rm HA}^{-{1\over2}}(L^{10}_{\rm HA})^{3\over2}\cr
L^5_{\rm IA}&=(L^{10}_{\rm HA})^{1\over2}\lambda_{\rm HA}^{1\over2},
}}
with the $K3$'s radii related as follows:
\eqn\radiii{R_{\rm IA}=R_{\rm HA}(L^{10}_{\rm HA})^{1\over2}
\lambda_{\rm HA}^{-{1\over2}}.}

\subsec{Geometrical Picture of the Dualities}

Taking into account all of the steps we have performed in constructing
the chain of dualities, it should now be clear that the organizing
geometry is that of a cylinder\horavawitten. Essentially, we
compactify M--theory on the cylinder given by
$S^1_{10}{\times}S^1_5/\IZ_2$, with length $L^{10}_{\rm M}$ for the ordinary
circle and length $L^5_{\rm M}$ for the line interval.

Shrinking $L^{10}_{\rm M}$ to zero sends us to the type~IA string theory limit
while shrinking $L^5_{\rm M}$ sends us to the $E_8{\times}E_8$ heterotic
limit. Once in the string theory limit,  $T$--duality
along the surviving direction of the cylinder takes us to type~IB string
theory in case ($x^{5}$), or $Spin(32)/\IZ_2$ heterotic string
theory in the other case ($x^{10}$).

The strong/weak duality exchange of the type~IB and $Spin(32)/\IZ_2$
heterotic string is simply the exchanges of the values of $L^5_{\rm M}$ and
$L^{10}_{\rm M}$, taking one from an infinitely thin line segment in one the
first case to an infinitely short cylinder ---the $x^{10}$ circle---
in the other.

\subsec{Origin of Heterotic/Heterotic Duality}

So far, we have not mentioned the extra special property this family
of vacua have, the property which gives rise to heterotic/heterotic
duality in six dimensions.

Its origin is clear once we realize that there is an alternative
route from the type~IB model to a heterotic model. Before going to
type~IA, we could perform the $T$--duality operation $T_{6789}$, which
has the effect of defining a very similar type~IB model, but on a $K3$
manifold obtained by inverting the radii of all of the circles in the
defining $T^4$. In fact, the model is invariant under this operation,
as can be seen by examining the orientifold group\ericjoe. 

Indeed, it is instructive to consider (separately) two interesting
D--string probes in the six uncompactified directions, one a
D1--brane, lying in the $(x^0,x^4)$ directions, and another a
D5$^\prime$--brane wrapped about the $K3$ giving a string lying in the
$(x^0, x^5)$ direction.

\vbox{
$$\vbox{\offinterlineskip
\hrule height 1.1pt
\halign{&\vrule width 1.1pt#
&\strut\quad#\hfil\quad&
\vrule width 1.1pt#
&\strut\quad#\hfil\quad&
\vrule#
&\strut\quad#\hfil\quad&
\vrule#
&\strut\quad#\hfil\quad&
\vrule#
&\strut\quad#\hfil\quad&
\vrule#
&\strut\quad#\hfil\quad&
\vrule#
&\strut\quad#\hfil\quad&
\vrule
\hskip 2pt
\vrule#
&\strut\quad#\hfil\quad&
\vrule#
&\strut\quad#\hfil\quad&
\vrule#
&\strut\quad#\hfil\quad&
\vrule#
&\strut\quad#\hfil\quad&
\vrule width 1.1pt#\cr
height3pt
&\omit&
&\omit&
&\omit&
&\omit&
&\omit&
&\omit&
&\omit&
&\omit&
&\omit&
&\omit&
&\omit&
\cr
&\hfil type&
&\hfil $x^0$&
&\hfil $x^1$&
&\hfil $x^2$&
&\hfil $x^3$&
&\hfil $x^4$&
&\hfil $x^5$&
&\hfil $x^6$&
&\hfil $x^7$&
&\hfil $x^8$&
&\hfil $x^9$&
\cr
height3pt
&\omit&
&\omit&
&\omit&
&\omit&
&\omit&
&\omit&
&\omit&
&\omit&
&\omit&
&\omit&
&\omit&
\cr
\noalign{\hrule height 1.1pt}
height3pt
&\omit&
&\omit&
&\omit&
&\omit&
&\omit&
&\omit&
&\omit&
&\omit&
&\omit&
&\omit&
&\omit&
\cr
&\hfil D1&
&\hfil --- &
&\hfil $\bullet$ &
&\hfil $\bullet$ &
&\hfil $\bullet$ &
&\hfil --- &
&\hfil $\bullet$ &
&\hfil $\bullet$ &
&\hfil $\bullet$ &
&\hfil $\bullet$ &
&\hfil $\bullet$ &
\cr
height3pt
&\omit&
&\omit&
&\omit&
&\omit&
&\omit&
&\omit&
&\omit&
&\omit&
&\omit&
&\omit&
&\omit&
\cr
\noalign{\hrule}
height3pt
&\omit&
&\omit&
&\omit&
&\omit&
&\omit&
&\omit&
&\omit&
&\omit&
&\omit&
&\omit&
&\omit&
\cr
&\hfil D5 (D1$^\prime$)&
&\hfil --- &
&\hfil $\bullet$ &
&\hfil $\bullet$ &
&\hfil $\bullet$ &
&\hfil $\bullet$ &
&\hfil --- &
&\hfil --- &
&\hfil --- &
&\hfil --- &
&\hfil --- &
\cr
height3pt
&\omit&
&\omit&
&\omit&
&\omit&
&\omit&
&\omit&
&\omit&
&\omit&
&\omit&
&\omit&
&\omit&
\cr
}\hrule height 1.1pt
}
$$
{\bf Table 3.} {\it Orientation of D--brane probes in the orientifold
model of Table 1 representing type~IB string theory on $K3$, giving
rise to two D--strings in six dimensions.} }

The operation $T_{6789}$ defines type~IB on the ``dual'' $K3$ which we
call $\widetilde{K3}$, with radius $1/R_{\rm IB}$.  $T_{6789}$ simply
turns the D5--brane into a D1--brane and {\it vice--versa}, therefore
exchanging our two types of D--string. Their tensions are easily
computed\berkoozi\ in terms of the type~IB and the $Spin(32)/\IZ_2$ heterotic
parameters, using eqn.\radii:
\eqn\tensionsi{\eqalign{{\cal T}_{\rm D1}&={1\over\lambda_{\rm IB}}=
{R^2_{\rm HB}\over R^2_{\rm IB}}\cr 
{\cal T}_{{\rm D1}^\prime}&=
{R^4_{\rm IB}\over\lambda_{\rm IB}}=R^2_{\rm HB}R^2_{\rm IB},}}
and are clearly exchanged under $T_{6789}$. 

In purely $Spin(32)/\IZ_2$ heterotic string paramters, their tensions
are (respectively)\berkoozi:
\eqn\tensionsi{\eqalign{{\cal T}_{\rm electric}&=
R^2_{\rm HB}\lambda^6_{\rm HB}\cr {\cal T}_{\rm magnetic}&={R^2_{\rm
HB}\over\lambda^6_{\rm HB}},}} where $\lambda^6_{\rm HB}=R^{-2}_{\rm
IB}$ is the six dimensional heterotic string coupling. We see that the
D1--brane becomes an electric ``fundamental'' heterotic string, being
light at weak heterotic coupling. Meanwhile the D1$^\prime$--brane
becomes a magnetic ``soliton'' string, being heavy at weak coupling.
Type~IB's $T_{6789}$--duality, which inverts $R_{\rm IB}$, maps here
to a strong/weak coupling duality.

Notice also that although we exchange two $K3$'s (with inversely
related volumes) in the type~IB model, the $K3$ which the heterotic
string is compactified on remains the same, as can be seen by using
eqn.\radii, with the $T_{6789}$--duality relations
\eqn\relations{T_{6789}:\biggl\{\eqalign{R_{\rm IB}&\to R^{-1}_{\rm IB}\cr
\lambda_{\rm IB}&\to\lambda_{\rm IB}/R^4_{\rm IB}}\biggr.,}
revealing that $R_{\rm HB}$ is invariant under the duality.

The orientations of the strings are trivially related to those in the
first type~IB model by a rotation by $\pi/2$, $O_{45}$, in the $(x^4,
x^5)$ plane.

\vbox{
$$\vbox{\offinterlineskip
\hrule height 1.1pt
\halign{&\vrule width 1.1pt#
&\strut\quad#\hfil\quad&
\vrule width 1.1pt#
&\strut\quad#\hfil\quad&
\vrule#
&\strut\quad#\hfil\quad&
\vrule#
&\strut\quad#\hfil\quad&
\vrule#
&\strut\quad#\hfil\quad&
\vrule#
&\strut\quad#\hfil\quad&
\vrule#
&\strut\quad#\hfil\quad&
\vrule
\hskip 2pt
\vrule#
&\strut\quad#\hfil\quad&
\vrule#
&\strut\quad#\hfil\quad&
\vrule#
&\strut\quad#\hfil\quad&
\vrule#
&\strut\quad#\hfil\quad&
\vrule width 1.1pt#\cr
height3pt
&\omit&
&\omit&
&\omit&
&\omit&
&\omit&
&\omit&
&\omit&
&\omit&
&\omit&
&\omit&
&\omit&
\cr
&\hfil type&
&\hfil $x^0$&
&\hfil $x^1$&
&\hfil $x^2$&
&\hfil $x^3$&
&\hfil $x^4$&
&\hfil $x^5$&
&\hfil $x^6$&
&\hfil $x^7$&
&\hfil $x^8$&
&\hfil $x^9$&
\cr
height3pt
&\omit&
&\omit&
&\omit&
&\omit&
&\omit&
&\omit&
&\omit&
&\omit&
&\omit&
&\omit&
&\omit&
\cr
\noalign{\hrule height 1.1pt}
height3pt
&\omit&
&\omit&
&\omit&
&\omit&
&\omit&
&\omit&
&\omit&
&\omit&
&\omit&
&\omit&
&\omit&
\cr
&\hfil D5 (D1$^\prime$)&
&\hfil --- &
&\hfil $\bullet$ &
&\hfil $\bullet$ &
&\hfil $\bullet$ &
&\hfil --- &
&\hfil $\bullet$ &
&\hfil --- &
&\hfil --- &
&\hfil --- &
&\hfil --- &
\cr
height3pt
&\omit&
&\omit&
&\omit&
&\omit&
&\omit&
&\omit&
&\omit&
&\omit&
&\omit&
&\omit&
&\omit&
\cr
\noalign{\hrule}
height3pt
&\omit&
&\omit&
&\omit&
&\omit&
&\omit&
&\omit&
&\omit&
&\omit&
&\omit&
&\omit&
&\omit&
\cr
&\hfil D1&
&\hfil --- &
&\hfil $\bullet$ &
&\hfil $\bullet$ &
&\hfil $\bullet$ &
&\hfil $\bullet$ &
&\hfil --- &
&\hfil $\bullet$ &
&\hfil $\bullet$ &
&\hfil $\bullet$ &
&\hfil $\bullet$ &
\cr
height3pt
&\omit&
&\omit&
&\omit&
&\omit&
&\omit&
&\omit&
&\omit&
&\omit&
&\omit&
&\omit&
&\omit&
\cr
}\hrule height 1.1pt
}
$$
{\bf Table 4.} {\it Orientation of D--brane probes of an orientifold
representing type~IB on $\widetilde{K3}$ (see Table~1), which is
$T_{6789}$ dual to the orientations of Table 3. These probes also give
rise to two D--strings in six dimensions.} }

Instead of doing the operation $T_{6789}$ we may proceed to our
type~IA model, using $T_5$--duality. We thus obtain type~IA on
$K3{\times}S^1_5/\IZ_2$. Here our two types of brane probes become a
D2--brane (with one leg stretched between the end of the $x^5$
interval) and a wrapped (on $K3$) D4--brane. (See Table~5)

\vbox{
$$\vbox{\offinterlineskip
\hrule height 1.1pt
\halign{&\vrule width 1.1pt#
&\strut\quad#\hfil\quad&
\vrule width 1.1pt#
&\strut\quad#\hfil\quad&
\vrule#
&\strut\quad#\hfil\quad&
\vrule#
&\strut\quad#\hfil\quad&
\vrule#
&\strut\quad#\hfil\quad&
\vrule#
&\strut\quad#\hfil\quad&
\vrule#
&\strut\quad#\hfil\quad&
\vrule
\hskip 2pt
\vrule#
&\strut\quad#\hfil\quad&
\vrule#
&\strut\quad#\hfil\quad&
\vrule#
&\strut\quad#\hfil\quad&
\vrule#
&\strut\quad#\hfil\quad&
\vrule width 1.1pt#\cr
height3pt
&\omit&
&\omit&
&\omit&
&\omit&
&\omit&
&\omit&
&\omit&
&\omit&
&\omit&
&\omit&
&\omit&
\cr
&\hfil type&
&\hfil $x^0$&
&\hfil $x^1$&
&\hfil $x^2$&
&\hfil $x^3$&
&\hfil $x^4$&
&\hfil $x^5$&
&\hfil $x^6$&
&\hfil $x^7$&
&\hfil $x^8$&
&\hfil $x^9$&
\cr
height3pt
&\omit&
&\omit&
&\omit&
&\omit&
&\omit&
&\omit&
&\omit&
&\omit&
&\omit&
&\omit&
&\omit&
\cr
\noalign{\hrule height 1.1pt}
height3pt
&\omit&
&\omit&
&\omit&
&\omit&
&\omit&
&\omit&
&\omit&
&\omit&
&\omit&
&\omit&
&\omit&
\cr
&\hfil D2&
&\hfil  ---  &
&\hfil  $\bullet$  &
&\hfil  $\bullet$  &
&\hfil  $\bullet$  &
&\hfil  --- &
&\hfil  [---]  &
&\hfil  $\bullet$ &
&\hfil  $\bullet$ &
&\hfil  $\bullet$ &
&\hfil  $\bullet$ &
\cr
height3pt
&\omit&
&\omit&
&\omit&
&\omit&
&\omit&
&\omit&
&\omit&
&\omit&
&\omit&
&\omit&
&\omit&
\cr
\noalign{\hrule}
height3pt
&\omit&
&\omit&
&\omit&
&\omit&
&\omit&
&\omit&
&\omit&
&\omit&
&\omit&
&\omit&
&\omit&
\cr
&\hfil D4&
&\hfil --- &
&\hfil $\bullet$ &
&\hfil $\bullet$ &
&\hfil $\bullet$ &
&\hfil $\bullet$ &
&\hfil $\bullet$ &
&\hfil --- &
&\hfil --- &
&\hfil --- &
&\hfil --- &
\cr
height3pt
&\omit&
&\omit&
&\omit&
&\omit&
&\omit&
&\omit&
&\omit&
&\omit&
&\omit&
&\omit&
&\omit&
\cr
}\hrule height 1.1pt
}
$$
{\bf Table 5.} {\it Two types of D--brane probe in the orientifold
model of Table 2 realizing type~IA on $K3{\times}S^1_5/\IZ_2$.} }

These two arrangements of branes (and the complete orientifold model)
are dual to each other under $T_{456789}$--duality, which inverts the
$K3$ and results in a model of type~IA on
$\widetilde{K3}{\times}S^1_4/\IZ_2$. See Table~6. (A final $\pi/2$ rotation
$O_{45}$ would trivially restore the identical orientation.) This
model is $T_4$--dual to the type~IB on $\widetilde{K3}$ model of
Table~4.

\vbox{
$$\vbox{\offinterlineskip
\hrule height 1.1pt
\halign{&\vrule width 1.1pt#
&\strut\quad#\hfil\quad&
\vrule width 1.1pt#
&\strut\quad#\hfil\quad&
\vrule#
&\strut\quad#\hfil\quad&
\vrule#
&\strut\quad#\hfil\quad&
\vrule#
&\strut\quad#\hfil\quad&
\vrule#
&\strut\quad#\hfil\quad&
\vrule#
&\strut\quad#\hfil\quad&
\vrule
\hskip 2pt
\vrule#
&\strut\quad#\hfil\quad&
\vrule#
&\strut\quad#\hfil\quad&
\vrule#
&\strut\quad#\hfil\quad&
\vrule#
&\strut\quad#\hfil\quad&
\vrule width 1.1pt#\cr
height3pt
&\omit&
&\omit&
&\omit&
&\omit&
&\omit&
&\omit&
&\omit&
&\omit&
&\omit&
&\omit&
&\omit&
\cr
&\hfil type&
&\hfil $x^0$&
&\hfil $x^1$&
&\hfil $x^2$&
&\hfil $x^3$&
&\hfil $x^4$&
&\hfil $x^5$&
&\hfil $x^6$&
&\hfil $x^7$&
&\hfil $x^8$&
&\hfil $x^9$&
\cr
height3pt
&\omit&
&\omit&
&\omit&
&\omit&
&\omit&
&\omit&
&\omit&
&\omit&
&\omit&
&\omit&
&\omit&
\cr
\noalign{\hrule height 1.1pt}
height3pt
&\omit&
&\omit&
&\omit&
&\omit&
&\omit&
&\omit&
&\omit&
&\omit&
&\omit&
&\omit&
&\omit&
\cr
&\hfil D4&
&\hfil --- &
&\hfil $\bullet$ &
&\hfil $\bullet$ &
&\hfil $\bullet$ &
&\hfil $\bullet$ &
&\hfil $\bullet$ &
&\hfil --- &
&\hfil --- &
&\hfil --- &
&\hfil --- &
\cr
height3pt
&\omit&
&\omit&
&\omit&
&\omit&
&\omit&
&\omit&
&\omit&
&\omit&
&\omit&
&\omit&
&\omit&
\cr
\noalign{\hrule}
height3pt
&\omit&
&\omit&
&\omit&
&\omit&
&\omit&
&\omit&
&\omit&
&\omit&
&\omit&
&\omit&
&\omit&
\cr 
&\hfil D2&
&\hfil  ---  &
&\hfil  $\bullet$  &
&\hfil  $\bullet$  &
&\hfil  $\bullet$  &
&\hfil  [---]  &
&\hfil  --- &
&\hfil  $\bullet$ &
&\hfil  $\bullet$ &
&\hfil  $\bullet$ &
&\hfil  $\bullet$ &
\cr 
height3pt
&\omit&
&\omit&
&\omit&
&\omit&
&\omit&
&\omit&
&\omit&
&\omit&
&\omit&
&\omit&
&\omit&
\cr
}\hrule height 1.1pt
}
$$
{\bf Table 6.} {\it D--brane probe arrangements $T_{456789}$--dual to
those in table 5. The background is an orientifold model of type~IA on
$\widetilde{K3}{\times}S^1_4/\IZ_2$.} }

Upon moving to strong type~IA coupling, we get two apparently distinct
M--theory models, one on $K3{\times}S^1_5/\IZ_2$ and the other with
$\widetilde{K3}{\times}S^1_4/\IZ_2$. Our D2--brane probe becomes an
M2--brane with one leg wrapped in the line interval~${\cal I}_5$
(${\cal I}_4$), and transverse to the $K3$ (or $\widetilde{K3}$). The
D4--brane becomes an M5--brane wrapped on the $K3$ (or
$\widetilde{K3}$) and transverse to the~${\cal I}_5$ (or~${\cal
I}_4$). See Tables~7 and~8.

\vbox{
$$\vbox{\offinterlineskip
\hrule height 1.1pt
\halign{&\vrule width 1.1pt#
&\strut\quad#\hfil\quad&
\vrule width 1.1pt#
&\strut\quad#\hfil\quad&
\vrule#
&\strut\quad#\hfil\quad&
\vrule#
&\strut\quad#\hfil\quad&
\vrule#
&\strut\quad#\hfil\quad&
\vrule#
&\strut\quad#\hfil\quad&
\vrule#
&\strut\quad#\hfil\quad&
\vrule
\hskip 2pt
\vrule#
&\strut\quad#\hfil\quad&
\vrule#
&\strut\quad#\hfil\quad&
\vrule#
&\strut\quad#\hfil\quad&
\vrule#
&\strut\quad#\hfil\quad&
\vrule
\hskip 2pt
\vrule#
&\strut\quad#\hfil\quad&
\vrule width 1.1pt#\cr
height3pt
&\omit&
&\omit&
&\omit&
&\omit&
&\omit&
&\omit&
&\omit&
&\omit&
&\omit&
&\omit&
&\omit&
&\omit&
\cr
&\hfil type&
&\hfil $x^0$&
&\hfil $x^1$&
&\hfil $x^2$&
&\hfil $x^3$&
&\hfil $x^4$&
&\hfil $x^5$&
&\hfil $x^6$&
&\hfil $x^7$&
&\hfil $x^8$&
&\hfil $x^9$&
&\hfil $x^{10}$&
\cr
height3pt
&\omit&
&\omit&
&\omit&
&\omit&
&\omit&
&\omit&
&\omit&
&\omit&
&\omit&
&\omit&
&\omit&
&\omit&
\cr
\noalign{\hrule height 1.1pt}
height3pt
&\omit&
&\omit&
&\omit&
&\omit&
&\omit&
&\omit&
&\omit&
&\omit&
&\omit&
&\omit&
&\omit&
&\omit&
\cr
&\hfil M2&
&\hfil --- &
&\hfil $\bullet$ &
&\hfil $\bullet$ &
&\hfil $\bullet$ &
&\hfil --- &
&\hfil [---]  &
&\hfil $\bullet$ &
&\hfil $\bullet$ &
&\hfil $\bullet$ &
&\hfil $\bullet$ &
&\hfil $\bullet$ &
\cr
height3pt
&\omit&
&\omit&
&\omit&
&\omit&
&\omit&
&\omit&
&\omit&
&\omit&
&\omit&
&\omit&
&\omit&
&\omit&
\cr
\noalign{\hrule}
height3pt
&\omit&
&\omit&
&\omit&
&\omit&
&\omit&
&\omit&
&\omit&
&\omit&
&\omit&
&\omit&
&\omit&
&\omit&
\cr
&\hfil M5&
&\hfil --- &
&\hfil $\bullet$ &
&\hfil $\bullet$ &
&\hfil $\bullet$ &
&\hfil $\bullet$ &
&\hfil $\bullet$ &
&\hfil --- &
&\hfil --- &
&\hfil --- &
&\hfil --- &
&\hfil --- &
\cr
height3pt
&\omit&
&\omit&
&\omit&
&\omit&
&\omit&
&\omit&
&\omit&
&\omit&
&\omit&
&\omit&
&\omit&
&\omit&
\cr
}\hrule height 1.1pt
}
$$
 {\bf Table 7.} {\it Two configurations of M--theory brane on
 $K3{\times}{\cal I}_5$.}  }

\vbox{
$$\vbox{\offinterlineskip
\hrule height 1.1pt
\halign{&\vrule width 1.1pt#
&\strut\quad#\hfil\quad&
\vrule width 1.1pt#
&\strut\quad#\hfil\quad&
\vrule#
&\strut\quad#\hfil\quad&
\vrule#
&\strut\quad#\hfil\quad&
\vrule#
&\strut\quad#\hfil\quad&
\vrule#
&\strut\quad#\hfil\quad&
\vrule#
&\strut\quad#\hfil\quad&
\vrule
\hskip 2pt
\vrule#
&\strut\quad#\hfil\quad&
\vrule#
&\strut\quad#\hfil\quad&
\vrule#
&\strut\quad#\hfil\quad&
\vrule#
&\strut\quad#\hfil\quad&
\vrule
\hskip 2pt
\vrule#
&\strut\quad#\hfil\quad&
\vrule width 1.1pt#\cr
height3pt
&\omit&
&\omit&
&\omit&
&\omit&
&\omit&
&\omit&
&\omit&
&\omit&
&\omit&
&\omit&
&\omit&
&\omit&
\cr
&\hfil type&
&\hfil $x^0$&
&\hfil $x^1$&
&\hfil $x^2$&
&\hfil $x^3$&
&\hfil $x^4$&
&\hfil $x^5$&
&\hfil $x^6$&
&\hfil $x^7$&
&\hfil $x^8$&
&\hfil $x^9$&
&\hfil $x^{10}$&
\cr
height3pt
&\omit&
&\omit&
&\omit&
&\omit&
&\omit&
&\omit&
&\omit&
&\omit&
&\omit&
&\omit&
&\omit&
&\omit&
\cr
\noalign{\hrule height 1.1pt}
height3pt
&\omit&
&\omit&
&\omit&
&\omit&
&\omit&
&\omit&
&\omit&
&\omit&
&\omit&
&\omit&
&\omit&
&\omit&
\cr 
&\hfil M5&
&\hfil --- &
&\hfil $\bullet$ &
&\hfil $\bullet$ &
&\hfil $\bullet$ &
&\hfil $\bullet$ &
&\hfil $\bullet$ &
&\hfil --- &
&\hfil --- &
&\hfil --- &
&\hfil --- &
&\hfil --- &
\cr
height3pt
&\omit&
&\omit&
&\omit&
&\omit&
&\omit&
&\omit&
&\omit&
&\omit&
&\omit&
&\omit&
&\omit&
&\omit&
\cr
\noalign{\hrule}
height3pt
&\omit&
&\omit&
&\omit&
&\omit&
&\omit&
&\omit&
&\omit&
&\omit&
&\omit&
&\omit&
&\omit&
&\omit&
\cr 
&\hfil M2&
&\hfil --- &
&\hfil $\bullet$ &
&\hfil $\bullet$ &
&\hfil $\bullet$ &
&\hfil [---]  &
&\hfil --- &
&\hfil $\bullet$ &
&\hfil $\bullet$ &
&\hfil $\bullet$ &
&\hfil $\bullet$ &
&\hfil $\bullet$ &
\cr
height3pt
&\omit&
&\omit&
&\omit&
&\omit&
&\omit&
&\omit&
&\omit&
&\omit&
&\omit&
&\omit&
&\omit&
&\omit&
\cr
}\hrule height 1.1pt
}
$$
 {\bf Table 8.} {\it Two configurations of M--theory brane on
 $\widetilde{K3}{\times}{\cal I}_4$.}  }

From either one of these M--theory models, we may proceed to recover
the heterotic string model by shrinking the line interval as discussed
before. In both cases, the M2--brane is seen to become the light,
electric, ``fundamental'' heterotic string (``F1--brane'') in the six
dimensions $(x^0, x^1, x^2, x^3, x^{4(5)}, x^{10})$. Meanwhile, the
M5--brane, wrapped on the $K3_{\rm M}$ ($\widetilde{K3}_{\rm M}$)
become a wrapped (on $K3_{\rm H}$ ) heterotic fivebrane soliton\nsfivebrane\
(``F5--brane'') yielding a heavy solitonic string in the six
dimensions. See Table~9.

\vbox{
$$\vbox{\offinterlineskip
\hrule height 1.1pt
\halign{&\vrule width 1.1pt#
&\strut\quad#\hfil\quad&
\vrule width 1.1pt#
&\strut\quad#\hfil\quad&
\vrule#
&\strut\quad#\hfil\quad&
\vrule#
&\strut\quad#\hfil\quad&
\vrule#
&\strut\quad#\hfil\quad&
\vrule#
&\strut\quad#\hfil\quad&
\vrule
\hskip 2pt
\vrule#
&\strut\quad#\hfil\quad&
\vrule#
&\strut\quad#\hfil\quad&
\vrule#
&\strut\quad#\hfil\quad&
\vrule#
&\strut\quad#\hfil\quad&
\vrule
\hskip 2pt
\vrule#
&\strut\quad#\hfil\quad&
\vrule width 1.1pt#\cr
height3pt
&\omit&
&\omit&
&\omit&
&\omit&
&\omit&
&\omit&
&\omit&
&\omit&
&\omit&
&\omit&
&\omit&
\cr
&\hfil type&
&\hfil $x^0$&
&\hfil $x^1$&
&\hfil $x^2$&
&\hfil $x^3$&
&\hfil $x^4$&
&\hfil $x^6$&
&\hfil $x^7$&
&\hfil $x^8$&
&\hfil $x^9$&
&\hfil $x^{10}$&
\cr
height3pt
&\omit&
&\omit&
&\omit&
&\omit&
&\omit&
&\omit&
&\omit&
&\omit&
&\omit&
&\omit&
&\omit&
\cr
\noalign{\hrule height 1.1pt}
height3pt
&\omit&
&\omit&
&\omit&
&\omit&
&\omit&
&\omit&
&\omit&
&\omit&
&\omit&
&\omit&
&\omit&
\cr
&\hfil F1 &
&\hfil  ---  &
&\hfil  $\bullet$  &
&\hfil  $\bullet$  &
&\hfil  $\bullet$  &
&\hfil  --- &
&\hfil  $\bullet$ &
&\hfil  $\bullet$ &
&\hfil  $\bullet$ &
&\hfil  $\bullet$ &
&\hfil  $\bullet$ &
\cr
height3pt
&\omit&
&\omit&
&\omit&
&\omit&
&\omit&
&\omit&
&\omit&
&\omit&
&\omit&
&\omit&
&\omit&
\cr
\noalign{\hrule}
height3pt
&\omit&
&\omit&
&\omit&
&\omit&
&\omit&
&\omit&
&\omit&
&\omit&
&\omit&
&\omit&
&\omit&
\cr
&\hfil F5 &
&\hfil --- &
&\hfil $\bullet$ &
&\hfil $\bullet$ &
&\hfil $\bullet$ &
&\hfil $\bullet$ &
&\hfil --- &
&\hfil --- &
&\hfil --- &
&\hfil --- &
&\hfil --- &
\cr
height3pt
&\omit&
&\omit&
&\omit&
&\omit&
&\omit&
&\omit&
&\omit&
&\omit&
&\omit&
&\omit&
&\omit&
\cr
}\hrule height 1.1pt
}
$$
{\bf Table 9.} {\it The orientation of the two branes (an F1--brane
and an F5--brane) which give rise to two dual $E_8{\times}E_8$
heterotic strings in six dimensions. One is a light ``fundamental''
string, while the other is a heavy solitonic string.} }

This heterotic model is compactified on $K3_{\rm H}$, with six
dimensional coupling $\lambda^6_{\rm H}{\sim}R^{-2}_{\rm
IB}$. Strong/weak coupling duality takes this to another model (with
coupling $1/\lambda^6_{\rm H}$) compactified on the same surface, with
an exchange of the F1-- and F5--branes. After a trivial relabeling of
the $x^5$ coordinate to $x^4$, we see that this is the model arising
from reduction of the other M--theory configuration (table~10):

\vbox{
$$\vbox{\offinterlineskip
\hrule height 1.1pt
\halign{&\vrule width 1.1pt#
&\strut\quad#\hfil\quad&
\vrule width 1.1pt#
&\strut\quad#\hfil\quad&
\vrule#
&\strut\quad#\hfil\quad&
\vrule#
&\strut\quad#\hfil\quad&
\vrule#
&\strut\quad#\hfil\quad&
\vrule#
&\strut\quad#\hfil\quad&
\vrule
\hskip 2pt
\vrule#
&\strut\quad#\hfil\quad&
\vrule#
&\strut\quad#\hfil\quad&
\vrule#
&\strut\quad#\hfil\quad&
\vrule#
&\strut\quad#\hfil\quad&
\vrule
\hskip 2pt
\vrule#
&\strut\quad#\hfil\quad&
\vrule width 1.1pt#\cr
height3pt
&\omit&
&\omit&
&\omit&
&\omit&
&\omit&
&\omit&
&\omit&
&\omit&
&\omit&
&\omit&
&\omit&
\cr
&\hfil type&
&\hfil $x^0$&
&\hfil $x^1$&
&\hfil $x^2$&
&\hfil $x^3$&
&\hfil $x^5$&
&\hfil $x^6$&
&\hfil $x^7$&
&\hfil $x^8$&
&\hfil $x^9$&
&\hfil $x^{10}$&
\cr
height3pt
&\omit&
&\omit&
&\omit&
&\omit&
&\omit&
&\omit&
&\omit&
&\omit&
&\omit&
&\omit&
&\omit&
\cr
\noalign{\hrule height 1.1pt}
height3pt
&\omit&
&\omit&
&\omit&
&\omit&
&\omit&
&\omit&
&\omit&
&\omit&
&\omit&
&\omit&
&\omit&
\cr 
&\hfil F5 &
&\hfil --- &
&\hfil $\bullet$ &
&\hfil $\bullet$ &
&\hfil $\bullet$ &
&\hfil $\bullet$ &
&\hfil --- &
&\hfil --- &
&\hfil --- &
&\hfil --- &
&\hfil --- &
\cr
height3pt
&\omit&
&\omit&
&\omit&
&\omit&
&\omit&
&\omit&
&\omit&
&\omit&
&\omit&
&\omit&
&\omit&
\cr
\noalign{\hrule}
height3pt
&\omit&
&\omit&
&\omit&
&\omit&
&\omit&
&\omit&
&\omit&
&\omit&
&\omit&
&\omit&
&\omit&
\cr 
&\hfil F1 &
&\hfil  ---  &
&\hfil  $\bullet$  &
&\hfil  $\bullet$  &
&\hfil  $\bullet$  &
&\hfil  --- &
&\hfil  $\bullet$ &
&\hfil  $\bullet$ &
&\hfil  $\bullet$ &
&\hfil  $\bullet$ &
&\hfil  $\bullet$ &
\cr 
height3pt
&\omit&
&\omit&
&\omit&
&\omit&
&\omit&
&\omit&
&\omit&
&\omit&
&\omit&
&\omit&
&\omit&
\cr
}\hrule height 1.1pt
}
$$
{\bf Table 10.} {\it Branes which give dual $E_8{\times}E_8$ heterotic
strings in six dimensions. } }

The two isomorphic tables above (9 and 10) are (by
construction)  respectively mapped directly under the
strong/weak coupling duality of type~IB and heterotic
strings\foot{Strictly, as discussed earlier, we must also perform a
$T_{10}$ duality operation to move from $E_8{\times}E_8$ to
$Spin(32)/\IZ_2$ heterotic.}\ to the type~IB models of tables 3 and 4,
where $x^5$ ($x^4$) is exchanged with $x^{10}$.

So the $T_{6789}$--duality map between the two type~IB models is
pulled back to the strong/weak coupling\foot{It could not have mapped
to a $T$--duality, as there is no such symmetry between the F1--brane
and the F5--brane.}\ heterotic/heterotic duality of Duff, Minasian and
Witten{\duff}, as pointed out in ref.\berkoozi. Here, we have traced
its action all the way through M--theory.

\subsec{Heterotic/Heterotic Duality as Eleven Dimensional 
Electromagnetic Duality}

Returning to the two M--theory models (tables 7 and 8) momentarily, it
is natural to ask about the nature of the map between them. In the
type~I theories it was simply $T$--duality at work, while in the
heterotic case it was six--dimensional electromagnetic strong/weak
coupling duality.  These must translate into something else in the
full M--theory.

The answer is that as eleven dimensional models, the two models are
electromagnetic duals of each other. It was suggested in ref.\duff\
that the operation of shrinking the interval~$\cal I$ with the
M2--brane wrapped on it was dual to another operation.  Performing an
eleven dimensional electromagnetic duality operation (with respect to
the potential $A^{(3)}$) involves exchanging the sources, replacing the
M2--brane with an M5--brane. Shrinking the M2--brane on~$\cal I$ is
then expected to be equivalent to wrapping the M5--brane on $K3$ and
shrinking the volume of the $K3$. This conjecture was partially
strengthened in ref.\duff\ by noting that precisely the (12,12)
arrangement of instantons was needed to allow an M5--brane wrapped
$K3$ to be shrunken completely.

However, there were geometric complications with making the
construction completely precise. In order to argue that the result is
again a heterotic string on $K3$, ref.\duff\ uses the intermediate
seven dimensional duality between M--theory on $K3$ and the heterotic
string on $T^3$, and then reduces to six dimensions on the interval
$\cal I$. This certainly gives a heterotic string in six dimensions,
but gives no complete geometrical understanding of how the transverse
$T^3{\times}{\cal I}$ really becomes a $K3$ again, much less the {\it
same} $K3$ on which the dual string is propagating.

The construction we have just performed in the previous subsection is
a means of making  the geometric proof complete. The key is that
precisely for this situation, the details of the electromagnetic
duality between the required configurations as eleven dimensional
models are {\it completely} implied by the $T_{456789}$ duality
between the two type~{\rm IA} models, with their accompanying brane
configurations depicted in tables~6 and~7.

The electromagnetic duality should be expected to work as follows: The
M2--brane couples to the potential $A^{(3)}$ electrically and carries
the basic unit of charge, while the M5--brane couples to it
magnetically, being the dual object to the M2--brane in eleven
dimensions.

In otherwise flat eleven dimensions with branes of infinite extent,
the operation of replacing~$A^{(3)}$'s field strength $F^{(4)}$ by its
dual $F^{(7)}$, while replacing M2--branes by M5--branes (and {\it
vice--versa}), would be a symmetry of the theory.

Here however, things are slightly more complicated. We have M--theory
on $K3_{\rm M}{\times}{\cal I}_5$, with sizes $R_{\rm M}$ and
$L^5_{\rm M}$, respectively. The M2--brane is an open M2--brane, as it
has boundaries on the ends of the interval~${\cal I}_{5}$. Part of
its volume is finite, with a scale set by $L^5_{\rm M}$. If
electromagnetic duality is going to work, we must conclude that the
world volume of the dual M5--brane must also have a finite part.  The
scale of that volume must be a monotonically increasing function of
$L^5_{\rm M}$, if we are to recover the uncompactified situation of
the previous paragraph in the limit.

This finite part of the M5--brane world volume is wrapped on a {\it
different} $K3$ from the one in the starting configuration, which we
can call $\widetilde{K3}_{\rm M}$. The radius $\widetilde{R}_{\rm M}$
must be an increasing function of $L^5_{\rm M}$. (It has to be a
different $K3$, as the other one must stay fixed while we reduce the
M2--brane--wrapped interval ${\cal I}_5$.)

Similarly, had we started with an M5--brane wrapped on the $K3_{\rm
M}$, with radius $R_{\rm M}$ we must conclude that the dual
configuration is an M2--brane wrapped on a line interval we might call
${\cal I}_4$, with length $L^4_{\rm M}$ an increasing function of
$R_{\rm M}$, and transverse to a {\it different}~$K3$ called
$\widetilde{K3}_{\rm M}$.

Precisely this situation is what we have in tables~7 and 8.  The
well--established $T_{456789}$--duality between these two
configurations tells us precisely how the eleven dimensional duality
must work. The parameters of one M--theory configuration are given by
the equations
\paramsii\ relating them to the parameters of the type~IA
configuration in table~5. Meanwhile, upon using the relations of
$T_{456789}$ duality between table~5 and table~6 (together with
$L^4_{\rm M}{=}\lambda^{-1/3}_{\rm IA}L^4_{\rm IA}$):
\eqn\tduality{\eqalign{R_{\rm IA}&\leftrightarrow
{1\over R_{\rm IA}}
\cr
L^{5(4)}_{\rm IA}&\leftrightarrow {1\over
L^{5(4)}_{IA}}\cr 
\lambda_{\rm IA}&\leftrightarrow 
{\lambda_{\rm IA}\over L^4_{\rm IA}L^5_{\rm IA}(R_{\rm IA})^4} }} we
can deduce the parameters of the dual M--theory configuration to be
related thus:
\eqn\relations{\eqalign{
\widetilde{R}_{\rm M}&=(L^5_{\rm M}L^4_{\rm M}R_{\rm M})^{1\over3}\cr
\widetilde{L}^{4(5)}_{\rm M}&=\left[(R_{\rm M})^4
{L^{5(4)}_{\rm M}\over(L^{4(5)}_{\rm M})^{2}}\right]^{1\over3}.  }} So
far these relations are apparently in line with our expectations above
that shrinking~${\cal I}_5$ is dual to shrinking $\widetilde{K3}_M$,
and also that shrinking the dual interval ${\cal I}_4$ is equivalent
to shrinking $K3_{\rm M}$. However, we must be careful. At face value,
they also seem to suggest that shrinking ${\cal I}_5$ results in the
shrinking of ${\cal I}_4$, which certainly does not seem correct.  

To set things aright, it is illuminating to factor out of the
expressions the quantity $V_{\rm M}{\equiv} L_{\rm M}^4L_{\rm
M}^5(R_{\rm M})^4$, the volume of the compact part $K3_{\rm M}{\times}{\cal
I}_5{\times}S^1_4$ of the M--theory compactification, whose image in
type~IA we $T_{456789}$--dualize. Doing this, we obtain:
\eqn\invert{\eqalign{
\widetilde{R}_{\rm M}&={V_{\rm M}^{1\over3}\over R_{\rm M}}\cr
\widetilde{L}^{4(5)}_{\rm M}&={V_{\rm M}^{1\over3}\over L^{4(5)}_{\rm M}}\cr
\hbox{\rm and also}\quad\widetilde{L}^{10}_{\rm M}&=
{V_{\rm M}^{-{2\over3}}\over{L}^{10}_{\rm M}}.  }}

These relations look like an M--theory analogue of type~IA's
$T_{456789}$--duality, with a volume factor entering in an unusual way
({\it c.f.} eqn.\tduality). Notice however that the dual volume is the
same as the original volume:
\eqn\volumes{\widetilde{V}_{\rm M}{\equiv}
\widetilde{L}_{\rm M}^4{L}_{\rm M}^5(\widetilde{R}_{\rm M})^4
={V^{1\over3}_{\rm M}\over L_{\rm M}^4}{V^{1\over3}_{\rm M}\over
L_{\rm M}^5}{V^{4\over3}_{\rm M}\over (R_{\rm M})^4}=V_{\rm M} ,} and
so the apparent length inversions in eqn.\invert\ are internal
rearrangements within the ``dualized'' surface, performed {\it at fixed
volume}. This is closer to what we should expect from the
electromagnetic duality, and less like a stringy $T$--duality.

We therefore interpret equation \invert\ as follows. First, we will
keep the volume $V_{\rm M}$ fixed while shrinking any part of $K3_{\rm
M}{\times}{\cal I}{\times}S^1$. Then, shrinking ${\cal I}_5$, we must
rescale the radius~$R_{\rm M}$ of the~$K3_{\rm M}$ to keep the volume
fixed.  In the dual M--theory configuration, this is equivalent to
(according to equation \invert) shrinking $\widetilde{R}_{\rm M}$, the
radius of the dual $\widetilde{K3}_{\rm M}$. The same reasoning may
be used to relate the shrinking of the dual interval ${\cal I}_4$ to
the shrinking of $K3_{\rm M}$ in the original geometry.

So we have sharpened the statement of Duff, Minasian and Witten
concerning the eleven dimensional origin of heteroitc/heterotic
duality: Wrapping an M2--brane on the interval~${\cal I}_5$ and
shrinking the interval to zero size to obtain a heterotic string
compactified on $K3_{\rm H}$ to six dimensions, is equivalent (by
electromagnetic duality) to wrapping an M5--brane on
$\widetilde{K3}_{\rm M}$ and shrinking the volume of the
$\widetilde{K3}_{\rm M}$ to zero to get a dual heterotic string
compactified on the {\it same} $K3_{\rm H}$ to six dimensions.

In summary, our main tool to study the details of the eleven
dimensional duality was to realize it as the image of T--duality
between two type~IA configurations.  Those configurations extend to
two apparently different M--theory compactifications under the strong
coupling map, and the duality map between the two M--theory
configurations follows from the existence of the T--duality map
composed with the (unique) strong/weak coupling map between of type~IA
and M--theory on a circle.

In this way, we also see that there is no mystery of how we recover 
$K3_{\rm H}$ for the resulting heterotic string  in the case of the
$K3_{\rm M}$--shrink--wrapping of the M5--brane, {\it and} we see that we
recover the correct $K3_{\rm H}$ as well.

DMW's result that there is no obstruction to shrinking the dual
M5--brane on a $K3$ when there is a (12,12) instanton compactification
directly translates into the statement that $T_{456789}$--duality
relates two type~IA configurations which can be taken to M--theory at
strong coupling, and thus in turn becomes eleven dimensional
electromagnetic duality in the limit.



\newsec{\bf Probing with D--Branes}
Let us now return to the study of the full details of the type~I
orientifolds model defined in tables~1 and~2. At the bottom of table~2
is shown the orientation of a D0--brane probe which we might introduce
into the model.

This is the starting point for the derivations of the various models
which we use to motivate parts of a matrix theory representation of
the $K3_{\rm M}{\times}{\cal I}$ compactification of M--theory.  To
derive a matrix theory we will use not just one D0--brane, but $N$ of
them. Following Banks, Fischler, Shenker and Susskind\bfss, and in the
notation of the previous section, the large $N$ limit will yield a
matrix theory representation of M--theory on $K3_{\rm M}{\times}{\cal
I}_5$ in the infinite momentum frame, where the infinite boost is in
the $x^{10}$ direction.

The logic of the procedures will be as follows: Considering first only
the local geometry for a moment, we see that we can derive (as the
effective D0--brane world--line theory) a model of~0+1 dimensional
quantum mechanics which encapsulates the physics of the probe moving
around in the background of the D8-- and D4--branes, the O8-- and
04--planes, and the $K3$. Given the strong coupling discussions in
section~2.3.2, in the matrix model (and large $N$) limit, such a 0+1
dimensional model should capture aspects of the M--theory physics of
small $E_8$ instantons on an ALE space, near an end of the
interval~${\cal I}_5$.

In moving on to consider the case where the compactness of the~${\cal
I}_5$ comes into play, we should expect that we must take into account
light strings interconnecting the $N$ D0--branes after first wrapping
this interval when it is small.  Operationally, (after a Fourier
transform in the matrix theory language) this gives us a 1+1
dimensional model, derivable as the physics of the $T_5$--dual
situation with $N$ $S^1_5$--wrapped D1--brane probes, oriented as at
the bottom of table~1.

This interlaces nicely with the fact that a D1--brane probe in type~IB
string theory is the heterotic string\edjoe: In the $L_5{\to}0$ limit,
the matrix theory is an uncompactified 1+1 dimensional model. This 1+1
dimensional matrix theory is the matrix theory of the heterotic string
on a local piece of $K3$, {\it i.e.,} near an ALE space, with (part
of) an $E_8$ instanton on it. In the strong coupling limit, which is
the IR limit of the model, we expect to find a conformal field theory
of this string, along the lines of Dijkgraaf, Verlinde and
Verlinde\dvv\ for the recovery of ten dimensional type~IIA strings
from matrix theory\foot{It may be interesting to see also the comments
at the end of ref.\robme, for some anticipation of this way of
realizing duality.}.

Furthermore, when we consider the next feature ---that the $K3$ is also
small and compact--- we must take into account light strings coming
from winding the $K3$. Again, we see that this is morally equivalent
to taking the $T_{6789}$--dual of the D1--branes defining the 1+1
dimensional model and working with the world--volume theory of the
dual probe, wrapped on a dual surface. In this way we see that we are
working with the world volume theory of $N$  D5$^\prime$--branes shown
at the bottom of table~1, and we therefore have (at least part of) a
definition in terms of a particular 5+1 dimensional quantum theory
compactified on the surface $\widetilde{K3}{\times}S_5^1$.

Notice that heterotic/heterotic duality is already manifest in this
matrix theory definition. Recall from section~2 that this duality is
generated by $T_{6789}$--duality, which in this context exchanges the
$N$ D1--branes with the $N$ D5$^\prime$--branes. 

As this theory is related in this way to the world volume of a
collection of $N$ D5$^\prime$--branes in type~IB string theory, we
suspect that it is related to the $(0,1)$  theory with
$Spin(32)/\IZ_2$ global symmetry\seiberg. We will thus be led to
examine the details of that expectation (in section~4), and
furthermore to consider what it means to compactify such a theory on
$\widetilde{K3}{\times}S_5^1$. Much later, we will speculate that
moving away from the special DMW/GP point, we should consider the
matrix theory definition of M--theory on $K3{\times}{\cal I}_5$ to be
given by the (0,1) $E_8{\times}E_8$ theory compactified on the dual
given by $\widetilde{K3}{\times}S^1_5$.

\subsec{\sl Probing with D1--Branes: The Model.}
As a starting point, we study the field content of the world--volume
theory of the $N$ D1--brane probes in the type~IB model. Returning to
the type~IA case with $N$ D0--branes will be easily performed by
(essentially) dimensional reduction.

We will refer to table~1 for the orientations of the GP model's branes
and the $K3$ (top part) and the $N$ D1--brane probes (bottom).

$\underline{\hbox{\it 3.1.1 Supersymmetry}}$

The D--brane orientations of the model (with D1--branes) breaks the Lorentz
group up as follows:

\eqn\lorentz{ 
 SO(1,9) \supset SO(1,1)^{05}\times SO(4)^{1234}\times SO(4)^{6789},}
where the superscripts denote the sub--spacetimes in which the
surviving factors act. Following refs.\refs{\wittenadhm,\douglasii}, we
may label the worldsheet fields according to how they transform under
the covering group (which acts as an R--symmetry of our final 1+1
dimensional model):
\eqn\cover{G=
[SU(2)^\prime\times \widetilde{SU(2)}^\prime]_{1234}\times
[SU(2)_R\times SU(2)_L]_{6789},} with doublet indices
$(A^\prime,\tilde{A}^\prime,A,Y)$, respectively.

We will place the $K3$ into the mix by embedding the reflection
symmetry $R_{6789}$, into the $SO(4)_{6789}$. 


Accordingly, the supercharges decompose under \lorentz\ first (due to
the D--strings) as
\eqn\susy{{\bf 16}={\bf 8}_++{\bf 8}_-} where $\pm$ subscripts
denote a chirality with respect to $SO(1,1)$, and furthermore (due to
the D5--branes and the $K3$) each $\bf 8$ decomposes into a pair of
$\bf 4$'s of the $SO(4)$'s.  As we will see in due course, the various
projections will pick out one of these spinors to carry the
supersymmetry on the world sheet, giving a $1{+}1$ dimensional system
with $(0,4)$ supersymmetry, as we might expect from traditional
heterotic string (on~$K3$) considerations, keeping in the back of our
minds that this is of course dual to that very system\foot{Notice that
this is the same amount of supersymmetry as possessed by a system of
D9-- D5-- and D1--branes\douglasi. The $K3$ does not break any more
supersymmetry than the D5--branes already do\ericjoe.}. 

The spectrum of massless fields in the model will produce a family of
fields on the world--volume, which has coordinates $(x^0, x^5)$. 
The supersymmetry algebra is of the form:
\eqn\susiealgebra{\{Q^{AA^\prime},Q^{BB^\prime}\}
=\epsilon^{AB}\epsilon^{A^\prime B^\prime} P_-,}
where $P_-{\equiv}-i\partial/\partial\sigma^-$, where
$\sigma^\pm{=}(x^0\pm x^5)/2$. Here, $\epsilon^{AB}$ and
$\epsilon^{A^\prime B^\prime}$ are the antisymmetric tensors of
$SU(2)$ and $SU(2)^\prime$, respectively, $A,B,A^\prime$ and
$B^\prime$ being doublet indices.

$\underline{\hbox{\it 3.1.2 The 1--1 Strings}}$ 

As there are 8 Dirichlet--Dirichlet (DD) directions, the
Neveu--Schwarz (NS) sector has zero point energy $-1/2$. The massless
excitations form  vectors and scalars in 2D, and are
created as follows. For the vectors, the Neumann--Neumann (NN)
directions give excitations:
\eqn\oneonenn{A^{0,5}(x^0,x^5):\quad \lambda_V\psi^{0,5}_{-{1\over2}} |0>,}
where $\lambda_V$ is an $2N{\times}2N$ Chan--Paton matrix, which
satisfies (at a fixed point of $R_{6789}$):
\eqn\projecti{\eqalign{\Omega:&
\quad \lambda_V\to -\gamma_\Omega^{\phantom{-}}\lambda^T_V\gamma_\Omega^{-1}\cr
R_{6789}:&\quad \lambda_V\to
\phantom{-}\gamma_R^{\phantom{-}}\lambda_V\gamma_R^{-1},}} where the
$\gamma_{\Omega(R)}^{\phantom{-}}$ matrices are $2N{\times}2N$
matrices chosen to represent the action of $\Omega$ and $R_{6789}$ on
$\lambda_V$. We will use the same type of basis as ref.\ericjoe. The
solution of the constraint equations \projecti\ results in $\lambda$
being generators of the group $U(N)$.

Away from a fixed point, the second constraint in eqn.\projecti\ does
not apply, and $R_{6789}$ merely relates a D--brane to its
image. Instead, only the orientation projection acts, giving gauge
group $SO(N)$. This is consistent with the fact that $\Omega^2$ has
the same sign for D1--branes as for D9--branes\ericjoe.

The excitations coming from the 8 DD directions split into two
parts. There are fields associated with excitations in the
$x^{1,2,3,4}$ directions, and fields associated with the $x^{6,7,8,9}$
directions:
\eqn\oneonedd{\eqalign{\phi(x^0,x^5): 
&\quad\lambda_\phi\psi^{1,2,3,4}_{-{1\over2}} |0>
\cr\psi(x^0,x^5):&\quad
\lambda_\psi\psi^{6,7,8,9}_{-{1\over2}} |0>,}} where
 $\lambda_{\phi(\psi)}$ are
$N{\times}N$ Chan--Paton matrices satisfying (at a fixed point):
\eqn\projectii{\eqalign{\Omega: & \quad
\biggl\{\quad \eqalign{\lambda_\phi&\to 
-\gamma_\Omega^{\phantom{-}}\lambda^T_\phi\gamma_\Omega^{-1}\cr
\lambda_\psi&\to \phantom{-}
\gamma_\Omega^{\phantom{-}}\lambda^T_\psi\gamma_\Omega^{-1}}\biggr.,\cr
\cr
R_{6789}:&\quad \biggl\{\quad \eqalign{\lambda_\phi&\to
\phantom{-}\gamma_R^{\phantom{-}}\lambda_\phi\gamma_R^{-1}\cr
\lambda_\psi&\to -\gamma_R^{\phantom{-}}\lambda_\psi\gamma_R^{-1}}\biggr. .}}
(Once again, away from a fixed point, $R_{6789}$ places no constraint,
and simply reflects.)

Now since $\lambda_\phi$ satisfies the same constraints as
$\lambda_V$, we see that the $\phi$ fields are a family of
four--component scalars transforming in the adjoint of $U(N)$, (or of
$SO(N)$ away from a fixed point), which we shall denote as
$b^{A^\prime\tilde{A}^\prime}$, to make contact with
refs.\refs{\wittenadhm,\douglasii}. Meanwhile, the $\psi$ fields are a
complex doublet of hypermultiplet fields transforming in the
antisymmetric representation of the gauge group $U(N)$ or of
$SO(N)$. We denote them $b^{AY}(x^0,x^5)$, in accordance with
refs.\refs{\wittenadhm,\douglasii}.


The fermionic states $\xi$ from the Ramond (R) sector (with zero point
energy 0, by definition) are built on the vacua formed by the zero
modes $\psi_0^i,\quad i{=}0,\ldots,9$. This gives the initial $\bf 16$
of the left hand side of eqn. \susy. The GSO projection acts on the
vacuum in this sector as:
\eqn\gso{(-1)^F=\Gamma^0\Gamma^1\ldots\Gamma^9,}
while as $\Omega$ acts as $-1$ on NN strings ({\it i.e.,} in the $(x^0,
x^5)$ directions), it is:
\eqn\omegga{\Omega=
\Gamma^1\Gamma^2\Gamma^3\Gamma^4\Gamma^6\Gamma^7\Gamma^8\Gamma^9,}
and as $R_{6789}$ is $-1$ in the $x^{6,7,8,9}$ directions it is:
\eqn\R{R_{6789}=-\Gamma^6\Gamma^7\Gamma^8\Gamma^9.}
So we have $(-1)^F\xi{=}\xi$ from the GSO projection, and with
$\Omega$, it simply correlates world sheet chirality with spacetime
chirality: $\Gamma^0\Gamma^5\xi_{\pm}=\pm\xi_\pm$, where $\xi_-$ is in
the ${\bf 8}_c$ of $SO(8)$ and~$\xi_+$ is in the~${\bf 8}_s$.

Now $\xi_-$ is further decomposed by $R_{6789}$ into $\xi^1_-$ and
$\xi^2_-$, where superscripts~1 and 2 denote the decomposition into
the (1234) sector and the (6789) sector, respectively.  So we have
that the four component fermion $\xi^1_-$ (hereafter called
$\psi_-^{A\tilde{A}^\prime}$) is the right--moving superpartner of the
four component scalar field $b^{A^\prime\tilde{A}^\prime}$, while
$\xi^2_-$ (called $\psi_-^{A^\prime Y}$) is the
right--moving superpartner of $b^{AY}$. The supersymmetry
transformations are:
\eqn\susie{\eqalign{
\delta b^{A^\prime\tilde{A}^\prime}&=i\epsilon_{AB}\eta^{A^\prime A}_{+} 
\psi_-^{B\tilde{A}^\prime}\cr
\delta b^{AY}&=i\epsilon_{A^\prime B^\prime}\eta^{AA^\prime}_+
\psi_-^{B^\prime Y}.
}} The $\psi$ carry the same $U(N)$ (or $SO(N)$) charges as their
bosonic superpartners in order to survive the $R_{6789}$ and $\Omega$
projections, thus ensuring that gauge symmetry respects supersymmetry.

Similarly, the field $\xi_+$ is decomposed under \lorentz\ into
$\xi^1_+$ (which we'll call $\psi_+^{A\tilde{A}^\prime}$) and
$\xi^2_+$ (called $\psi_+^{A^\prime Y}$).  Formally, these fields are
left--moving ``superpartners'' of the gauge field $A^\mu$. We have
only $(0,4)$ supersymmetry, and so no {\sl linearly} realized
left--moving supersymmetry transformations, but there are non--linear
ones. As $A^\mu$ is not dynamical in two dimensions, the detailed form
of these transformations will not concern us here\foot{In the case of
a single D1--brane, $A^\mu$ and $\xi_+$ would be projected out by
acting with $\Omega$, leaving us with just the right--moving states
$\xi_-$ as superpartners of the $b$--fields, as in standard heterotic
lore\edjoe. In this case, we have Chan Paton factors $\lambda_V$, on
which a representation of $\Omega$ can act to supply us with a
correctly transforming surviving state.}.

$\underline{\hbox{\it 3.1.3 The 1--9 Strings}}$

There are 8 Dirichlet--Neumann (DN) coordinates, giving ground state
energy 1/2, and so there are no massless states arising in the NS
sector. The R sector excitations come from the NN $(x^0,x^5)$ system
giving just two ground states. In this sector, the 
GSO projection is simply $(-1)^F{=}\Gamma^0\Gamma^5$, which picks the
left--moving field\edjoe.

Assuming (for now) that we are in the configuration where we have
maximal gauge group $U(16)_9$ from the D9--branes, we have
left--moving fermions $\lambda_+$ in the $\bf{(N,16)}$, the first
part of the current algebra fermions of the heterotic string. We
denote a component by $\lambda_+^M$, where $M$ is a D9--brane index,
whenever we need to explicity show the index structure under the
global symmetry of the D9--brane gauge group.

(We shall see a little later that we have not completely determined
the 1--9 spectrum. There is another sector which arises which will
play a crucial role in the proceedings.)

\bigskip
\bigskip
\bigskip

$\underline{\hbox{\it 3.1.4 The 1--5 Strings}}$

There are four DN coordinates, and four DD coordinates giving the NS
sector a zero point energy of 0, with excitations coming from integer
modes in the $1234$ directions, giving a four component boson. This is
sector 1, in the notation above. The R sector also has zero point
energy of zero, with excitations coming from the 6789 directions,
giving a four component fermion $\chi$. This is sector 2, as above.

The GSO projections in either sector are:
\eqn\gsoagain{\eqalign{
(-1)^F_1&=\Gamma^0\Gamma^1\Gamma^2\Gamma^3\Gamma^4\Gamma^5\cr
(-1)^F_2&=\Gamma^0\Gamma^5\Gamma^6\Gamma^7\Gamma^8\Gamma^9, }} which,
upon application, reduce us to two bosonic states $\phi^{A^\prime}$ in
sector 2, and decomposes the spinor $\chi$ in sector 1 into left and
right moving two component spinors, $\chi_-^A$ and $\chi_+^Y$,
respectively. We see that $\chi_-^A$ is the right--moving superpartner
of $\phi^{A^\prime}$. 

Assuming also that we have chosen the configuration where all of the
D5--branes are coincident at one fixed point, we have therefore a
supermultiplet in the $\bf{(N,16)}$, with components~$(\phi^{A^\prime
m},\chi_-^{Am})$:
\eqn\susiei{\delta\phi^{A^\prime m}=i\epsilon_{AB}\eta^{A^\prime A}_{+}
\chi_-^{Bm}.} 
Here, $m$ is a D5--brane group theory index.

Also, (with components $\chi_+^{Ym}$), $\chi_+^{Y}$ transforms in the
$\bf{(N,16)}$. They are the rest of the current algebra fermions of
the heterotic string, which from this point of view carry any
non--perturbative global symmetry arising due to the D5--branes'
behaviour as small instantons.

$\underline{\hbox{\it 3.1.5 The 9--9, 5--5 and 9--5 Strings}}$

Crucial to the whole discussion is that fact that these fields will
generically appear as couplings in the 1+1 dimensional theory. 

From the 5--5 sector there are four--component scalar couplings,
(descendants of the $x^{6,7,8,9}$ hypermultiplets) transforming in the
${\bf 120}{+}\overline{\bf 120}$, which we call $X^{AY}_{mn}$,
matching the notation of ref.\douglasii.

There are similar fields in the 9--9 sector and we denote these couplings
$Y^{AY}_{MN}$. 

Meanwhile, the 9--5 sector produces a $\bf{(16,16)}$, denoted
$h^{Am}_M$, with $m$ and $M$ showing off its choices in D5-- and
D9--brane group theory.

Now that we have named these fields, we can display the supersymmetry
transformation relating them to the left moving fields:
\eqn\susieii{\eqalign{\delta\lambda_+^M&=\eta_+^{AA^\prime} C^M_{AA^\prime}\cr
\delta\chi_+^{Ym}&=\eta_+^{AA^\prime} C^{Ym}_{AA^\prime},}}
where:
\eqn\cform{\eqalign{C^M_{AA^\prime}&=h_{A}^{Mm}\phi_{A^\prime m}\cr
C^{Ym}_{AA^\prime}&=\phi_{A^\prime}^n(X^{Ym}_{An}-b_A^{Y}\delta_{n}^m),
}}

These precise transformations allow us to write the non--trivial part
of the $(0,4)$ supersymmetric 1+1 dimensional Lagrangian containing
the Yukawa couplings and the potential:

\eqn\lagrange{\eqalign{{\cal L}_{\rm tot}={\cal L}_{\rm kinetic}
-{i\over4}\int d^2\!\sigma\biggl[ &\lambda_+^M 
\left(
\epsilon^{BD}{\partial C^M_{BB^\prime} \over\partial b^{DY}}
\psi_-^{B^\prime Y}+
\epsilon^{B^\prime D^\prime}{\partial C^M_{B B^\prime}\over 
\partial\phi^{D^\prime m}}
\chi_-^{Bm}
\right)\biggr.\cr
\biggl.+&\chi_+^{Ym} 
\left(
\epsilon^{BD}{\partial C^{Ym}_{BB^\prime} \over\partial b^{DY}}
\psi_-^{B^\prime Y}+
\epsilon^{B^\prime D^\prime}{\partial C^{Ym}_{B B^\prime}\over 
\partial\phi^{D^\prime m}}
\chi_-^{Bm}
\right)\biggr.\cr
\biggl.+&{1\over2}
\epsilon^{AB}\epsilon^{A^\prime B^\prime}\left(C^M_{AA^\prime}
C^M_{BB^\prime}+C^{Ym}_{AA^\prime}C^{Ym}_{BB^\prime}\right)\biggr].
}}

This was derived in ref.\wittenadhm\ as the most general $(0,4)$
supersymmetric Lagrangian with these types of multiplets, providing
that the $C$ satisfy the condition:
\eqn\hyperkahler{C^M_{AA^\prime}
C^M_{BB^\prime}+C^{Ym}_{AA^\prime}C^{Ym}_{BB^\prime}+ C^M_{BA^\prime}
C^M_{AB^\prime}+C^{Ym}_{BA^\prime}C^{Ym}_{AB^\prime}=0,} which they
do\douglasii. ${\cal L}_{\rm kinetic}$ contains the usual kinetic terms
for all of the fields, and the required terms which complete them
into gauge invariant terms.

It should be of some concern that the 9--9 hypermultiplet fields
$Y^{AY}_{MN}$ have not entered into the story in a non--trivial way.
We see their role in what follows:

$\underline{\hbox{\it 3.1.6 Two Puzzles, and Their Solution}}$

Running a little ahead of the story for a minute, let us anticipate
what this model should do for us thus far. We have written down the
physics of the (multi--) D1--brane probe (slightly generalizing
ref.\refs{\douglasi,\douglasii}), and we expect (following
refs.\refs{\douglasii,\wittenadhm,\douglasmoore}) that the conditions
\hyperkahler\ will restrict us to the moduli space of vacua of the 1+1
dimensional theory which is isomorphic to the moduli space of
instantons. In other words, equations \hyperkahler\ are the ADHM data,
and finishing the search for the gauge inequivalent vacua by setting
the potential to zero and gauge fixing will compute a hyperK\"ahler
quotient for us. This makes contact with the fact\edsmall\ that the
D5--branes\foot{More precisely, a quartet of them\ericjoe. See also
section 2 in this paper.}\ are instantons in the D9--brane gauge
fields, dual to heterotic instantons.

This anticipation is correct. However, our studies in section~2 should
suggest that something is not quite right. Indeed, the D5--branes are
correctly to be seen as instantons in the model, but they are not the
only objects playing that role.  As there are only 8 full instantons
to be found in the D5--brane sector, the other 16 needed for a
consistent $K3$ compactification are to be found elsewhere: There is a
single instanton living at the core of each fixed point of $R_{6789}$.

So even in the complete absence of D5--branes, it must be that the
D1--brane probes should be able to detect instantons, by finding them
in the moduli space of their world--sheet sigma model. In the model
that we have above (eqn.\lagrange), if we crudely mimic the absence of
D5--branes by setting all of the 1--5, 5--5 and 9--5 fields to
zero\foot{To be less crude, we would actually just send the masses of
the 1--5 fields to infinity, telling us that the D5--branes are
infinitely far away.}, we are left with a trivial model with not
enough structure to give us ---after the hyperK\"ahler quotient---
anything other than either the trivial flat space (over--endowed with
a hyperK\"ahler structure) when we are away from a fixed point, or the
(resolved; see the next sub--subsection) ALE space.

So somehow things need to be modified in order that the D1--branes see
an instanton hiding in the ALE space. This is our first puzzle.

This is not unrelated to the apparent absence of a role for the 9--9
fields $Y^{AY}_{MN}$. If they are to be non--trivially involved, they
must somehow enter the lagrangian in a $(0,4)$ respecting way. The
only way to do so is to enter in equations of the form
\cform. However, an examination of the index structure shows that
there is nothing for $Y^{AY}_{MN}$ to couple to in order to enter
consistently. 

Persevering, an examination of the $\lambda_+^M$ supersymmetry
transformations \susieii\ which it must enter finds the source of the
problem: As it stands the supersymmetry transformations are of the
form 1-9{=}1-5$\cdot$5-9 or 1-5{=}1-5$\cdot$5-5, showing the fusion of
D1--, D5--, and D9-- index structure. There is no way to form such an
equation with this structure using $Y^{AY}_{MN}$ and the fields we
already have: We need a 1--9 field with indices ${A^\prime, M}$. This
constitutes our second puzzle.

If both problems find their solution in a common place it must be that
something happens at the fixed points of $R_{6789}$. This is precisely
at the heart of the first problem, as we know that away from the
fixed points, the system should be identical to that which has gone
before\refs{\douglasii,\wittenadhm}\ for the 1--5--9 system in flat
space.

Turning therefore back to our computation of the spectrum of the 1--9
sector we see that there is indeed something extra at a fixed
point. For a D1--brane sitting at a fixed point, there is a wholly new
sector of massless strings arising in the 1--9 sector.  These new
strings are special ({\it c.f.,} section 3.1.3) in that they have
their endpoints fixed in the $x^{6,7,8,9}$ directions.

As the literature does not seem to elaborate on such strings, let us
be explicit in showing where they come from: It is analogous to the
case of the extra massless 1--1 strings which arise when a D1--brane
alights upon a fixed point. First consider, in the covering space of
the orbifold, the D1--brane at some position ${\bf X}{=}{\bf X}_0$
away from the fixed point which is (say) at ${\bf X}{=}0$. The
massless 1--9 strings arise entirely from strings beginning on the
D1--brane and ending on the nearby D9--branes (which are of course
everywhere) at ${\bf X}_0$. These strings will also contribute to the
spectrum when the D1--brane is at the fixed point, and are considered
in section~3.1.3. 

There are however, 1--9 strings which stretch from the D1--brane to
the mirror image position on the D9--branes, $-{\bf X}_0$ (near the
mirror partner of the D1--brane, which is not relevant here). These
are stretched, and do not contribute to the massless spectrum. On the
orbifolded space, we must think of these strings as stretching from
the D1--brane at ${\bf X}_0$ to the fixed point ${\bf X}{=}0$, and
back to the vicinity of the D1--brane. Although it is back where it
started, due to the reflection, it is stretched to length $2{\bf
X}_0$, and looped through the fixed point. The figure shows both
points of view schematically (we have let the strings touch the
D9--branes slightly away from the D1--brane, for clarity):

\medskip
\centerline{\epsfsize1.5in\epsfbox{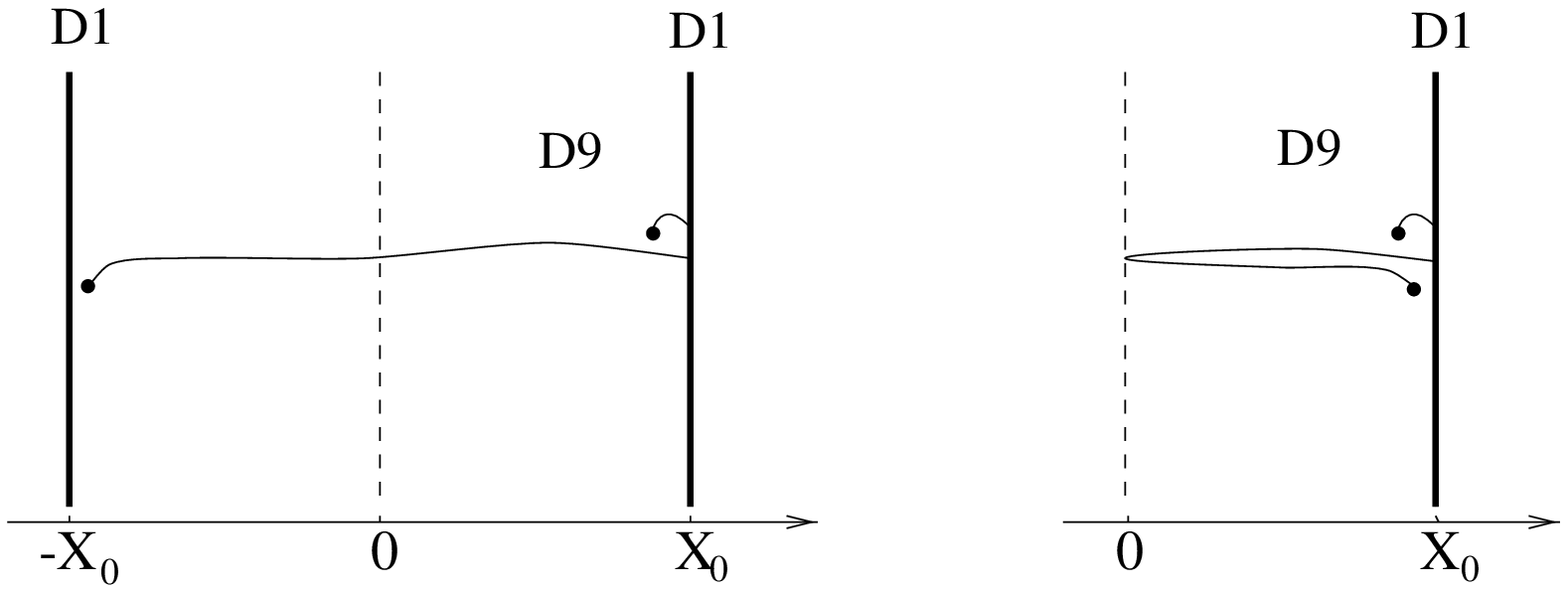}}
\medskip

As the D1--brane approaches the fixed point, these looped strings
become of zero length and therefore contribute to the massless
spectrum {\it in addition to the ordinary 1--9 strings}. The crucial
difference between these ``twisted 1--9 strings'' and the garden
variety is that they are trapped at the fixed point {\it i.e.,} in the
$x^{6,7,8,9}$ directions.  So instead of 8 DN coordinates, as computed
previously in section~3.1.3, giving a zero point energy of 1/2, there
are 4 DN coordinates $(x^{1,2,3,4})$ and 4 DD coordinates
$(x^{6,7,8,9})$, shifting the zero point energy to $-1/2$. Now we can
excite massless states in both the NS and R sector, just as computed
for the 1--5 sector previously\foot{Indeed, it is as if we have
introduced a whole new type of five--brane which is trapped at the
fixed point. We can think of this as an explicit orbifold realization
of the phenomenon of wrapping a D9--brane on $K3$ to get a
five--brane. This new five--brane is different from ordinary
D5--branes, though. It couples to the closed string twist fields,
which is another reason why it is stuck at the fixed points. The
cylinder and M\"obius strip diagrams of refs.\refs{\ericjoe.\ericmeI}\
which have twisted sector fields propagating in the closed string
channel, may be thought of as having these five--branes attached to
their boundaries.}.

In this way, we get three new fields for the 1--9 sector when we are
at a fixed point: the supermultiplet $(\rho^{A^\prime
M},\zeta_-^{AM})$, with supersymmetry transformations:
\eqn\susieiii{\delta\rho^{A^\prime M}
=i\epsilon_{AB}\eta^{A^\prime A}_{+}\zeta_-^{BM},} and the left mover
$\zeta^{YM}_+$. ($M$ is a D9--brane index.)

{\it This is precisely what we need to solve our problem.} At a fixed
point, we have in addition to the term in the Lagrangian involving
$\lambda^M_+$, we have a term
\eqn\replacement{\zeta_+^{YM} 
\left(
\epsilon^{BD}{\partial C^{YM}_{BB^\prime} \over\partial b^{DY}}
\psi_-^{B^\prime Y}+
\epsilon^{B^\prime D^\prime}{\partial C^{YM}_{B B^\prime}\over 
\partial\phi^{D^\prime M}}
\zeta_-^{BM}
\right),}
where
\eqn\cformii{C^{YM}_{AA^\prime}
=\rho_{A^\prime}^N(Y^{YM}_{AN}-b^{Y}_A\delta_N^M),}
and 
\eqn\susieiii{\delta\zeta_+^{YM}=\eta_+^{AA^\prime} C^{YM}_{AA^\prime},} 
which has a 1-9{=}1-9$\cdot$9-9 structure matching the
1-5=1-5$\cdot$5-5 structure of the second equation in \susieii.  We
complete the modification of our Lagrangian by adding obvious terms
involving $C^{YM}_{AA^\prime}$ to eqn.\hyperkahler.

This modification has a number of virtues, in solving our problem:

\item\item {\it (i)} It is the unique 
way that we can modify the 1--9 sector of the lagrangian and still
keep the supersymmetry structure.
\item\item{\it (ii)} We see how the
fixed points break D1--D5 translational symmetry {\it via} the inclusion of
the couplings \cformii. With the couplings \cform\ alone, there is
translational symmetry which should only be appropriate away from the
fixed points.
\item\item{\it (iii)} We see now that even
in the absence of all D5--brane fields, the D9--branes supply
couplings which satisfy the same form of (ADHM) specification
\hyperkahler\ that the D5--brane fields did. Therefore, the D1--branes
will correctly see instantons in the D9--branes, {\sl but only at the
fixed points !}

Let us now return to our story.

$\underline{\hbox{\it 3.1.7 The Closed Strings}}$

No less crucial are the couplings which descend from the
hypermultiplets fields arising in the closed string sector of the
orbifold. There are twenty such four--component scalar couplings, four
controlling the overall shape of the $K3$, while the other sixteen are
twisted sector fields, coming one per fixed point of $R_{6789}$.

A fixed point hypermultiplet's four components split into a
triplet~$\zzeta$ of NS--NS fields which transform as a vector of the
$SU(2)_R$ group, plus a R--R scalar $\phi^{(0)}$ which is a singlet.
As they are closed string fields, they have no group theory charges
from the gauge symmetry~(D1), or global symmetry (D5, D9) carried by
the branes.

They couple into the sigma model as follows. The potential for the
1--1 sector fields $b^{AY}$ arises in an ${\cal N}{=}4$ 1+1
dimensional D--term. In accord with the fact that we have an $SU(2)_R$
symmetry, we can write it in terms of a vector $\bf{\mmu}$. 

Using the notation presented in a similar context in ref.\robme, the
hypermultiplet field $b^{AY}$ is explicitly related to the quaternion:
\eqn\hyperquat{\Psi=\pmatrix{\psi^{1}_H&-\psi^{2\dagger}_H\cr
\psi^{2}_H&\phantom{-}\psi^{1\dagger}_H}=b,}
which acted on by $SU(2)_L$ and $SU(2)_R$ in the defining
representation in the obvious way.
We can then write the Lie algebra valued ``moment map'' vector as
\eqn\nodterms{\mmu^a\equiv{\rm Tr}
\left[\lambda^a_V\cdot\left\{\Psi^{1\dagger}_H {\bf\Sigma}\Psi^1_H
+\Psi^{2\dagger}_H {\bf\Sigma}\Psi^2_H\right\}
\right]}
where 
\eqn\quatern{\Psi^{1\dagger}_H=\left(\psi^{1}_H,-\psi^{2\dagger}_H\right)
\quad{\rm and}\quad
\Psi^{2\dagger}_H=\left(\psi^{2}_H,\psi^{1\dagger}_H\right) } 
are the natural $SU(2)_R$ doublets appearing in the quaternionic form
\hyperquat. The three 
components, $\sigma^i$, of ${\bf\Sigma}$ are the Pauli matrices acting
on the $SU(2)_R$ doublet space. The $\lambda^a_V$ are generators of
the gauge group under which the hypermultiplets are charged.

This vector enters into the Lagrangian coupling to an auxiliary field
$D$. The closed string fields $\zzeta$ also couple to the D--auxiliary
field, a term arising for each available $U(1)$ factor of the gauge
group. In this case, there is one such factor. After integrating $D$
out, we learn that $\bf\mmu$ and $\zzeta$ couple in Fayet--Illiopoulos
fashion
\eqn\dterm{\int d^2\!\sigma \left({\bf \mmu}^i-\zzeta\right)^2,} 
where $\mmu^i$ means the restriction of $\mmu$ in \nodterms\ to the
$U(1)$ subgroup. The non--Abelian parts of $\mmu$ simply give D--terms
with no FI term contributing\foot{That such couplings appear in the
physics of D--branes was demonstrated explicitly in calculations in
ref.\douglasmoore.}$^{,}$\foot{See refs.\refs{\joetensor,\robme} for
more details of how the coupling works in this context of D1--brane
probes.}. These D--terms are also an important part of the
hyperK\"ahler quotient, of course and should not be ignored. 

Away from the fixed point, the gauge group is $SO(N)$, and there is no
such coupling allowed. This is perfectly acceptable because the closed
string twisted sector fields are of course not present in this case.

This coupling is intimately related to the mechanism which is
responsible for breaking the $U(16)$'s to $SU(16)$'s (see
ref.\refs{\berkoozi,\douglasmoore}). In the 5+1 dimensional theory,
while the triplet~$\zzeta$ is involved in this FI term, the final
member of the supermultiplet,~$\phi^{(0)}$, has Chern--Simons
couplings to a six--form ``anomaly polynomial'' $X^{(6)}$, producing a
counterterm in the Lagrangian which allows the anomalous $U(1)$ to be
canceled in a generalized\sagnottii\ Green--Schwarz fashion.  To
complete the job, anomalous gauge transformations are given to
~$\phi^{(0)}$ which induce a mass term for the $U(1)$, breaking
$U(16)$ to $SU(16)$, in a manner generalizing the
Dine--Seiberg--Witten\dine\ mechanism.

\subsec{\sl Probing with D1--Branes: The Geometry}

$\underline{\hbox{\sl 3.2.1 Instantons, ALE Spaces, D--Flatness,
 HyperK\"ahler
Quotients and  ADHM Data.}}$

At the end of the day, we discover the background fields that the
D1--brane probe sees by simply finding the moduli space of possible
values which the fields can take, given that we must preserve
$(0,4)$ supersymmetry. This moduli space of vacua, given by $V{=}0$, where $V$
is the potential, should of course only include (world--sheet) gauge
inequivalent field configurations, if we are not to over specify the problem.

A shorthand phrase for this procedure is to perform the
``hyperK\"ahler quotient\hitchinetal''. Indeed, as first pointed out
in refs.\refs{\douglasii,\douglasmoore}, the physics of D--branes as
probes (with this amount of supersymmetry) is isomorphic to certain
classic mathematical work by Atiyah, Drinfeld, Hitchin and Manin
(ADHM)\refs{\ADHM}, by Kronheimer\refs{\kronheimer}, and by Kronheimer
and Nakajima\refs{\kronheimernakajima}.

This work is extremely relevant to us:

\item\item{\it (1)} The ADHM construction of instantons 
essentially constructs the instanton connections as 
\eqn\connectioni{A_\mu= U^\dagger\partial_\mu U,}
where $U$ is defined by
\eqn\define{\eqalign{{\cal D}^\dagger & U=0\cr
 U^\dagger & U=\II_2.}}  The operator $\cal D$ is constructed\foot{A
most enlightening paper where much of this is made explicit is
ref.\italiansi.}\ out of what might be called the ``ADHM data'', and
when it obeys
\eqn\obey{{\cal D}^\dagger {\cal D}=\Delta\otimes\II_2,}
where $\Delta$ is an invertible Hermitian constant matrix, the field
strength $F_{\mu\nu}$ of $A_\mu$ in \connectioni, is self--dual. ($U$
and $\cal D$ have quaternion values, in the $2{\times}2$ defining
representation of $SU(2)_R$, and $\II_2$ is the corresponding identity
matrix.)

These conditions translate directly into a series of equations for the
ADHM data. These equations are precisely the equations \hyperkahler\
ensuring $(0,4)$ supersymmetry of our particular sigma model,
\lagrange. As first pointed out in ref.\refs{\wittenadhm}, $V{=}0$ for
the sigma model (and removing gauge redundancy) gives a space of
solutions which is isomorphic to those of ADHM.

In particular, restricting to the massless modes, one finds that the
kinetic terms for the current algebra fermions are shifted to\wittenadhm:
\eqn\newkinetic{{i\over 2}\int d^2\!\sigma\sum_{i,j}\biggl\{
\lambda_{+i}^M\left(\delta_{ij}\partial_-+\partial_-
b^{\mu}A_{\mu ij}\right)\lambda_{+j}^N
\biggr\},}
where we have used the $x^6,x^7,x^8,x^9$ spacetime index $\mu$ on our 1--1
field $b^{AY}$ instead of the indices $(A,Y)$, for clarity.

That the ADMH conditions are the conditions \hyperkahler\ on 5--5 and
9--5 fields (couplings here) was shown explicitly in
ref.\refs{\douglasii}.

\item\item{\it (ii)} Kronheimer's construction of ALE
instantons is also relevant to us.  In ref.\refs{\kronheimer}, it was
shown that one can construct the ALE spaces using a hyperK\"ahler
quotient. The technique shows how to construct the full family of
spaces as functions of a set of deformation parameters
$\{\zzeta\}$. When $\{\zzeta{=}0\}$, one has the singular
spaces~$\IR^4/\Gamma$, where $\Gamma$ is any discrete subgroup of
$SU(2)$ while non--zero $\zzeta$ gives their smooth deformations
into ALE gravitational instantons.  ``ALE'' means that they are flat
at infinity, but not Euclidean, due to identifications by $\Gamma$ at
infinity. As $\Gamma$ has an A--D--E classification\mackay, so does
the family of ALE spaces\refs{\hitchin}.

In the case in hand, we have $\Gamma{=}\{1,R_{6789}\}{=}\IZ_2$, and
the resolved space is called the Eguchi--Hanson\refs{\eguchihanson}
space, controlled by a single deformation parameter $a$:
\eqn\eguchi{ds^2=\left(1-\left({a\over r}\right)^4\right)^{-1}dr^2
 +{r^4\over 4}(\sigma^2_x+\sigma^2_y)+{r^2\over 4}
\left(1-\left({a\over r}\right)^4\right)\sigma^2_z.}
Here, $\sigma_i$ are the familiar $SU(2)_R$--invariant one--forms of
$SU(2)$. The metric \eguchi\ is shown in terms of these to make its
$SU(2)_R$ invariance manifest. In Euler coordinates $(\phi, \theta,
\psi)$, we have:
\eqn\thesigma{\eqalign{\sigma_z&=d\psi+\cos\theta d\phi\cr
\sigma_y&=-\cos\psi d\theta-\sin\psi\sin\theta d\phi\cr
\sigma_x&=\sin\psi d \theta-\cos\psi\sin\theta d\phi\cr
{\rm giving}\,\,\,\sigma_x^2+\sigma_y^2&=d\theta^2+\sin^2\theta
d\phi^2.}}

It turns out that the physics of D--branes probing orbifold
singularities $\IR^4/\Gamma$ is isomorphic to Kronheimer's
construction, as shown explicitly for the A--series in
refs.\refs{\douglasmoore,\joetensor}\ and for the D and E series in
ref.\refs{\robme}\foot{See also the appendix of ref.\italiansi\ for the
explicit example of the derivation of the metric \eguchi\ with this
method.}. The data specifying the quotient is phrased in terms of a
``moment map'' $\mmu$ and some parameters $\zzeta$ (called the ``level
set''). In performing the hyperK\"ahler quotient, the equation
\eqn\blowup{\mmu^i-\zzeta{=}0}
 must be satisfied. This is what we  do (among other things, like
setting the other D--terms to zero) in setting $V{=}0$, when we
recognize that one of the terms in $V$ is given by the equation
\dterm. As mentioned in section~3.1.7, the moment map $\mmu$ is simply
a combination of 1--1 hypermultiplet fields $b^{AY}$, while the
deformation parameters $\zzeta$ are the NS--NS blowup modes coming
from the closed string twisted sectors.

\item\item{\it (iii)} Kronheimer and Nakajima constructed gauge
instantons on the ALE spaces by combining the techniques mentioned
above in {\it (i)} and {\it (ii)}. The self dual connections
$A$ on the ALE space are constructed as 
\eqn\connectionii{A_\mu= U^\dagger\nabla_\mu U,}
where $U$ satisfies similar equations as before in terms of an object
$\cal D$ which this time contains both ADHM type data and moment map
type data\foot{The ``covariant'' derivative $\nabla$ used in
\connectionii\ is defined with a certain connection $A^{\cal T}$ which is
an Abelian connection on the ``tautological bundle'' on the ALE
space. Simply put, the ``tautological bundle'' is a certain bundle on
an ALE space which comes essentially for free.  See point {\sl (iv)}
below, for ${\cal A}^T$'s crucial role in the context of this paper.}.
The conditions on $\cal D$ for self--duality are turn out to be of the
same form as equations
\hyperkahler, but deformed by~$\mmu$. 
The rest of the hyperK\"ahler quotient ($V{=}0$, etc) includes imposing
the blowup condition \blowup\ to make the ALE base space.

In our case, the equations which we have written immediately before
and after (and including) \lagrange\ are for the situation where we
are away from a fixed point. At a fixed point, we must use the
additional multiplets and coupling for the 1--9 sector as discussed in
section~3.1.6. The modified conditions for $(0,4)$ supersymmetry are,
in the basis we have chosen, of the same form as \hyperkahler, but
with extra terms added to include $C^{YM}_{AA^\prime}$. The
deformation parameter $\zzeta{=}\mmu^i$ is implicit, as $\mmu$ is part
of the potential for $b^{AY}$, given in terms of the $C$'s.

$\underline{\hbox{\sl 3.2.2 The Moduli Space of GP Models and Probe
Models}}$

So far, we have constructed everything about the D--probe model at the
highly symmetric point in the moduli space of GP models where we have
gauge symmetry $U(16)_9{\times}U(16)_5$ gauge symmetry. This comes
from having no Wilson lines for the D9--branes and all of the 32
D5--branes on one fixed point. This gauge symmetry acts as a global
symmetry of the 1+1 dimensional probe model, as the all of the fields
(except the 1--1 fields) are charged under it, and all of the
important couplings (except those descending from the closed string
sector) as well.

There is a whole family of models in the GP class, and hence by
extension, a whole family of D--probe models.  In general, there can
be an even number of D5--branes at any fixed point. For $m_I$ such
pairs at the $I$th fixed point, there is enhanced gauge symmetry
$U(m_I)_5$ contributed to the gauge group.  There are two 5--5
hypermultiplets $X^{AY}_{mn}$ in the ${1\over2}{\bf m}_I({\bf m}_I-1)$
dimensional antisymmetric representation, and a 9--5 hypermultiplet
$h^{A}_{Mn}$ in the $({\bf 16},{\bf m}_I)$.

For $n_J^\prime$ D5--branes at a non fixed point $J$, the gauge group
is $USp(2n_J^\prime)$, where $n_J^\prime$ must be a multiple of four,
the basic dynamical unit away from a fixed point. There is a 5--5
hyper $X^{AY}_{mn}$ in the ${1\over2}{\bf n}_J^\prime({\bf
n}_J^\prime-1)$ antisymmetric representation and a 9--5 hyper
$h^{A}_{Mn}$ in the $({\bf 16},{\bf n}_J^\prime)$. This is as it
should be for type~IB D5--branes in open space, where they should
behave like small $Spin(32)/\IZ_2$ instantons\edsmall.

The pattern just described is reflected in the structure of the
patterns of breaking which can occur with D9--brane Wilson lines, and
therefore the possible spectrum of gauge groups and hypermultiplets in
the 9--9 sector is identical.

This structure is inherited by the 1+1 dimensional probe model, the
1--9 fields $\lambda_+^M$, $\rho^{A^\prime M}$, and $\zeta^{AM}_\pm$,
and 1--5 fields $\phi^{A^\prime m}$ and $\chi^{Am}_\pm$, all
tranform in the fundamentals of the relevant gauge groups. The
9--9, 9--5 and 5--5 couplings are the hypermultiplets with charges
given above.

So we see that the various branches of the moduli space includes the
physics which we expected to see:
\item\item{\it (a)} D5--branes are instantons
in the D9--brane gauge fields. Away from the fixed points (where 4
D5--branes make a single instanton), the physics is simply that of
$Spin(32)/\IZ_2$ instantons, as pointed out in ref.\edsmall. The
expectation values of the 9--5 fields control the overall size of the
instanton and when their expectation values are zero we have enhanced
gauage symmetry $USp(2){\equiv}SU(2)$ for each isolated single
instanton and $USp(2k)$ when $k$ of them are coincident. Here $k$
is~8, at most because the 32 D5--branes are forced to move in groups
of four, as can be deduced from the allowed pattern of Higgs--ing in
the 5+1 dimensional model, which translates into an R--symmetry
restriction in the sigma model here.

\item\item{\it (b)} On the fixed points, D5--branes are 
also instantons, but are now instantons on the resolved space. The
small instanton limit there gives enhanced gauge symmetry $U(k)$
for~$k$ coincident {\sl half}--instantons (two D5--branes) on the
fixed point.

Note however that:

\item\item{\it (c)} In the absence of D5--branes in the vicinity, and 
we are just probing the fixed point, our ADHM instanton data is not
nearly as complicated as that which specifies the instantons which
D5--branes construct. This is because of the absence of an analogue of
the 9--5 hypermultiplet fields $h^{A}_{Mm}$. The vacuum expectation
values of the $h$ control the {\sl size} of the non--Abelian D5--brane
instantons. At the fixed point, when we the 1--9 fields $(\rho,\zeta)$
and their couplings in the sigma model, there is no analogue of the
$h$ fields to give a size modulus to whatever instantons the
D1--branes will see coming purely from the D9--branes and the fixed
point. So those instantons will be of a special type, geometrically,
having no size deformation.

\item\item{\it (d)} These special instantons will have no size deformations.
More correctly, they will have no size parameter independently of the
geometry of the resolved fixed point. In other words, their size is
set by the same parameter which sets the size of the blowup (see next
section). Intuitively, this singles out an (essentially) unique family
of instantons associated to the class of spaces which the fixed points
resolve to. These ``ALE'' spaces all possess, due to purely
geometrical considerations, a $U(1)$ monopole connection $A^{\cal T}$
which is naturally associated to the space\foot{They are connections
on the ``tautological bundle'' over the ALE space $X$. The first Chern
class of their field strengths $F$ decomposes under $\Gamma$ into a
basis for the $H^2(X,\IR)$ cohomology of ALE spaces, and the resulting
intersection matrix turns out to be isomorphic to the adjacency matrix
of the familiar A--D--E Dynkin diagrams. This in turn defines a family
of intersecting two--spheres which is the minimal resolution of the
orbifold singularity. But this is one beautiful story we must
regretfully neglect to tell in full.}. $A^{\cal T}$ is essentially a
monopole field, which is to say that it is an Abelian instanton. In
the case (most relevant here) of the Eguchi--Hanson space \eguchi, to
which our fixed points resolve, the $U(1)$ instanton connection is
\eqn\connectioniii{A^{\cal T}=-{a^2\over r^2}(d\psi+\cos\theta d\phi).}

We see that the single parameter $a$ controls the size of the
Eguchi--Hanson space \eguchi\ {\sl and} the (trivial) scale of the
Abelian instanton \connectioniii. This is awfully similar to what we
expect from the above comments (c) about the instanton which the
D1--branes should see living on the fixed point, and in the absence of
D5--branes.

Intuitively, this is clearly the instanton we
want\foot{Mathematically, a proof must work as follows: In order to
show that in the absence of 5--9 type couplings, the resulting
instantons are simply the canonical Abelian ones, it must be that the
matrix $U$ in eqn.\connectionii\ turns out to be a constant. Then the
solution $A{=}A^{\cal T}$ results, given the form of the covariant
derivative $\nabla{=}\partial{+}A^{\cal T}$.  After consulting
\kronheimernakajima, one sees that the 9--9 hypermultiplets are valued
in the endomorphisms $End(W)$ of a certain vector space $W$, while the
5--9 hypermultiplets are in the homomorphisms $Hom(V,W)$ from $W$ to
another vector space $V$. Roughly, the operator ${\cal D}^\dagger$
turns out to be a special map between spaces further defined by taking
certain combinations of $W,V$ and $\cal T$, the tautological bundle.
If $U$ is a constant it must mean (given \define) that the kernel of
the operator ${\cal D}^\dagger$, acting as a map, is trivial.  When
there is no involvement of the 5--9 fields (and hence no involvement
of structures in $Hom(V,W)$), it must be that the structure of the
linear maps that~${\cal D}^\dagger$ represents simplifies
considerably, resulting in the required simple result for $U$. The
resulting moduli of the reulting instantons must have dimension
zero. It would be interesting to complete that proof.}. What we have
done by discovering the required form of the sigma model at the fixed
point, is to take the Abelian instanton --- which is guaranteed to be
present for geometrical reasons --- and embed it into the D9--brane
gauge group. That final step of embedding this instanton, which is
otherwise just surplus to requirements, is not gauranteed at the
outset: we have to actually perform it in constructing the consistent
theory\foot{Contrast this with the case of the type~II theories ({\it
e.g.} ref.\robme\ in the same situation where the Abelian connections
on the ALE spaces are certainly there, but play no role in a stringy
gauge bundle.}. We did so by allowing the 9--9 fields to communicate
properly to the 1--1 sector via the new 1--9 fields we discovered.

Hence, we have made contact with section 4.1 of ref.\berkoozi. In the
orbifold limit, as we circumnavigate the fixed points we know that the
1--9 fields undergo a monodromy represented by a 32$\times$32
matrix~$M$ given by \eqn\M{M=\pmatrix{0&-I\cr I&0}} where $I$ is the
16$\times$16 identity matrix. This is inherited directly from the
choices made in ref.\ericjoe. Normalizing the Abelian instanton
$A^{\cal T}$ such that its monodromy about infinity matches $M$, gives
it $Spin(32)/\IZ_2$ instanton number 1, and ensures that it is an
instanton without ``vector structure'', which is to say that it does
not obey Dirac quantization in vector representations of $SO(32)$.

The surviving subgroup of $Spin(32)/\IZ_2$ after this embedding is the
$U(16)$ which we know that we have from the outset. We have therefore
seen from the worldsheet point of view just how the D1--branes see the
embedding of the Abelian instanton into the parent gauge group. 

$\underline{\hbox{\sl 3.2.3 The Special Point}}$

As mentioned and studied considerably in section~2, there is a certain
special point in the moduli space of GP models which is of great
interest. By extension, this point will also define a special 1+1
dimensional sigma model in the class we have built here.

This special point is the placement of two D5--branes on each of the
sixteen fixed points in the orbifold $K3$. The model is connected to
the situation where there is no gauge group. Considering the six
dimensional anomaly equation:
\eqn\anomaly{n_H-n_V=244+29 n_T,}
where $n_H, n_V$ and $n_T$ represent the number of hyper--, vector--
and tensor-- multiplets, respectively, this means that there are 244
hypermultiplets from the open and closed string sectors, as there are
no extra tensor multiplets in the spectrum beyond the standard one on
the supergravity multiplet\foot{See, however
refs.\refs{\ericmeI,\atish}\ for orientifold models similar to the GP
models which have extra tensors.}.

Let us see how the various hypermultiplet contributions arise at this
special point. Placing a single D5--brane pair at each of the sixteen
fixed points (and turning on appropriate Wilson lines on the torus) we
obtain gauge group $U(1)_9^{16}{\times}U(1)_9^{16}$. At this stage
$n_V{=}32$ and therefore $n_H{=}276$. By examining the details of the
GP spectrum given in section~3.2.2, we see that there are no 5--5 or
9--9 fields remaining, but there are 16 5--9 fields coming from each
of the sixteen fixed points making 256 9--5 fields. The remaining 20
are the $K3$ deformation hypermultiplets from the closed string
sector, 16 of them coming from the fixed points as twisted sector
fields.  As pointed out in ref.\berkoozi, 16 of the $U(1)$'s give rise
to anomalies which are canceled at one loop by a generalization of
the Green--Schwarz mechanism\refs{\sagnotti}\ which simultaneously
lifts the gauge symmetries by a generalization of the
Dine--Seiberg--Witten mechanism\refs{\dine}. 16 R--R components of the
closed string twist fields are involved in this process, their 16
NS--NS triplet partners, $\zzeta_i$, remaining to control the blowup
of the $K3$, as discussed in sections~3.1.7 and 3.2.1. Giving vacuum
expectation values to these will therefore remove the 16 anomalous
$U(1)$'s.  The remaining 16 $U(1)$'s can be Higgsed away, using
up a 9--5 field at each fixed point\foot{In the $T_{6789}$--model, the
5--5 $U(1)$'s are Higgs'd and the 9--9 $U(1)$'s are
Dine--Seiberg--Witten'd.}.

In the end, therefore, we see that the 244 hypermultiplets of the
spectrum of the special GP point are made up of 4 closed string
hypers, controlling the global geometry of the $K3$, and 15 9--5
fields coming from each fixed point. (The size of the vacuum
expectation values of the 16 closed string twist fields are still
adjustable in order to control the $K3$ shape, of course.)  These open
string hypermultiplets are also pure geometry: They are coordinates on
the Higgs branch (of the six dimensional field theory) of the $SO(32)$
(half--) instanton moduli space, representing the deformations away
from the zero size limit. As there are no 5--5 fields remaining, there
is no branch representing the motion of the point--like
half--instantons; they are stuck on the fixed points and all they can
do is grow and shrink.  There are no moduli for the Abelian instanton
of the 9--9 sector living at the fixed point, as we have seen
explicitly from equations
\eguchi\ and \connectioniii, and the discussion in section 3.2.2.

This is also confirmed by the index theorem computations in
ref.\berkoozi, where it was computed that on the manifold $K3$, the
dimension of the moduli space ${\cal M}_k$ of $k$ $Spin(32)/\IZ_2$
instantons which break the gauge group completely, with corrections
included to take into account the possibility of fractional
instantons, is
\eqn\modulispacei{{\rm dim}\,{\cal M}_k=120(k-1).}
From this, ref.\berkoozi\ saw that for $k{=}1$, the case of the
instanton inherent to the fixed point, we get the answer zero,
confirming that there are no moduli for that instanton.

Here, we see that this formula confirms everything else we have said
above, since with the half--instanton included, we have 3/2 of an
instanton at each point. Putting this into the formula, we get that
there is a 60 dimensional moduli space associated with this
point. Dividing by four to get the amount of hypermultiplets, this
corresponds to the 15 hypermultiplets which we found above.

In this way, we understand --- from many complementary points of view
--- just what the one dimensional probe of $N$ coincident D1--branes
sees as it moves around the interior of the~$K3$ of our
compactification. In particular, there is $3/2$ of an instanton stuck
at each fixed point of the orbifold $K3$.

\subsec{Probing with D0--branes and D5--branes.}

$\underline{\hbox{\sl 3.3.1 D0--brane probes}}$

Now we are in a position to consider the $N$ D0--brane probe 0+1
dimensional theory. There is not much to do, in terms of constructing
a Lagrangian, as it is obtained by simply dimensionally reducing the
one we have constructed in the previous subsection.

Thinking of this dimensional reduction as being performed on the $x^5$
circle, we must take into account the possibility of introducing
Wilson lines for the gauge field $A_\mu$ in the process. This results
in an extra massless parameter $X^5$ in the quantum mechanics
representing the position of the D0--branes along the dual $x^5$
direction, which is the line interval $\cal I$.  Simultaneously, from
the point of view of the D9-- and D5--branes' six dimensional gauge
fields, there is a reduction to five dimensions with a Wilson line.

For the D9--branes, the particular Wilson line we are interested in is
the one which places the resulting D8--branes (in the type~IIA
picture) symmetrically at the ends of the interval~${\cal I}$.
Meanwhile, for the D5--branes, we choose Wilson lines which place the
32 D4--branes in the configuration discussed in section~2.3.2. The $N$
D0--brane probe therefore sees 1 D4--brane plus 1 O4--plane at each
fixed point of the $K3$ at each end of the interval, equivalent to 3/4
of an instanton. 

(Perhaps less confusingly, we can equivalently think of this as the
same $K3$ as in the type~IB model, but now every point on it has the
content of a line interval $\cal I$. In this way, we see that there is
still the 3/2 instanton at each fixed point, but it further subdivided
into two parts placed at each end of ${\cal I}$.)

$\underline{\hbox{\sl 3.3.2 D5--brane probes}}$

A six dimensional field theory may be obtained from our 1+1
dimensional probe models by $T_{6789}$--duality. This turns our family
of $N$ coincident D1--brane probes into a family of $N$ coincident
D5$^\prime$--brane probes, in the notation of section~2 (see
Table~2). They are wrapped around the $K3$ of type~IB, and give rise
to a dual family of D1--brane probes. By the self--duality of the
model, these dual D1--probes have the same physics of the D1--brane
probes we discussed in section~3.1 and~3.2. They inherited this
physics from the wrapping of the D5$^\prime$--branes, and hence we
indirectly know some of the physics of the compactified $N$
D5$^\prime$--branes.

We can say a few things directly about the 5+1 dimensional
world--volume theory of the wrapped branes which will be useful in a
later discussion.

The background D5-- and D9--branes produce the same family of global
symmetries as they did in the 1+1 dimensional model, supplying charges
for the 5$^\prime$--5 and 5$^\prime$--9 fields. The 5--5, 9--9 and 5--9
fields enter again as parameters.

We have the usual ${\cal N}{=}1$ supersymmetric Yang--Mills terms in
the Lagrangian, in terms of gauge potential $A^\mu$, arising in the
5$^\prime$--5$^\prime$ sector from $(x^0,x^5,x^6,x^7,x^8,x^9)$
excitations, and hypermultiplets coming from the
5$^\prime$--5$^\prime$ sector with excitations in the
$(x^1,x^2,x^3,x^4)$ directions. The gauge group is $U(N)$, the same as
for the 1+1 dimensional model at a fixed point. To get the sector
where it is $SO(N)$, we simply introduce the Wilson lines which are
dual to moving the D1--brane probes away from them as described in
subsection~3.2.

Crucially, we also have terms of the form
\eqn\terms{\mu_5\int d^6\!y\,
 H^{(7)}\cdot H^{(7)} + A^{(2)}\wedge\left(R\wedge R+ F\wedge
F\right),} where $R$ is the Ricci two--form of $K3$ and $F$ is the
field strength of the gauge field $A^\mu$ $(\mu{=}\{0,5,6,7,8,9\})$
. $H^{(7)}$ is the field strength of the R--R potential $A^{(6)}$ (to
which the D5--branes couple ``electrically'' with charge $\mu_5$) and
$A^{(2)}$ is its ten dimensional dual.

Satisfying the resulting field equations requires that due to the
presence of the 16 sources of curvature ($K3$'s fixed points), we have
to have 16 gauge instanton sources. These are of course, string--like
objects in the world volume theory, which are known to be
D1--branes\refs{\douglasi}, objects which are sources of $A^{(2)}$.

We can say precisely what the nature of these 16 D1--branes living in
the D5--branes must be.  First of all, they are identical in
structure, as the 16 fixed points are all of the same type.  They are
necessarily transverse to the $K3$ (and hence the coordinates where the
gauge instanton they carry is located), which puts them in the
$(x^0,x^5)$ directions. Furthermore, they must have precisely the same
global symmetries on their world--sheet action as the D5--branes'
world--volume possesses. {\sl This identifies them with exactly the
family of D1--branes we studied in sections~3.1 and~3.2.}

The crucial lesson of this discussion, which we will revisit later on,
is as follows: There is a moduli space of $K3$ compactifications of
the 5+1 dimensional theory associated with the $N$
D5$^\prime$--branes, which encodes the data of the $T_{6789}$--dual
$K3$ compactification of the type~IB theory.  We start with $N$
D5$^\prime$--branes in flat space, which has a $U(N)$ gauge symmetry
and a $Spin(32)/\IZ_2$ global symmetry. We then compactify them on a
$K3$. The data associated to compactifying type~IB on the dual $K3$ is
encoded in the world--volume theory by the patterns of breakings of
the global symmetry and the pattern of induced couplings which
arise. A family of stringy states are seen to exist in the model. The
family of strings which can arise is also parameterized by the
patterns of breakings of the global symmetries, and the couplings
which can arise in the 1+1 dimensional world--sheet theory. We have
characterized such string sigma models in sections 3.1 and 3.2.

\newsec{\bf Some (Partial) Matrix Theory Proposals}

We are now in a position to suggest a role for the D--brane theories
we have been discussing in the context of Matrix theory. Let us do
this step by step:

\subsec{\sl Matrix Theory of M--theory on $X{\times}{\cal I}$.}

The 1+1 dimensional models which we constructed in section~3 can be
used as the starting point for a definition of a matrix theory
representation of M--theory compactified on the Eguchi--Hanson space
$X$ times a line interval ${\cal I}_5$, of length $L^5_{\rm M}$.  By
equations~\paramsii\ and~\paramsiii, we can deduce precisely what
string theory lengths, $L^5_{\rm IA}$ and $L^5_{\rm IB}$, we must use
to define the model.  In particular, the model is should be defined by
placing our 1+1 dimensional model ---at the special point--- on a
circle of radius $L^5_{\rm IB}$ and taking the large $N$ limit. As $N$
is correlated (by $T_5$--duality) to D0--brane number, we expect that
this model ---when accompanied by the large $\lambda_{\rm
IA}{=}R^{10}_{\rm M}$ (in 11D metric) limit--- should give a matrix
theory definition of the compactification in the infinite momentum
frame. We expect that the problems noticed at finite $N$ for Matrix
theories on such surfaces in ref.\douglasooguri\ will disappear at
large $N$, as suggested for example in ref.\fischler. This expectation
should be accompanied by some hope also, as the physics is rather less
constrained by supersymmetry than before\bfss: we only have eight
supercharges.

\subsec{\sl Matrix Theory of the $E_8{\times}E_8$ Heterotic String 
on $X{\times}{\cal I}$.}

One interesting limit of the previous compactification is of course
the $L^5_{\rm M}{\to}0$ limit where we should recover the
$E_8{\times}E_8$ heterotic string compactified on the Eguchi--Hanson
space~$X$, with 3/2 of an $E_8$ instanton\foot{Matrix theories of
$E_8$ instantons have been presented in the literature
lately\matrixeight, in work independent of the present paper.}.

In the context of our matrix model, this limit is defined by starting
with the large $(N,R^{10}_{\rm M})$ limit of our 1+1 dimensional model
--- at the special point --- of section~3. It is the decompactified
$x^5$ limit of the model above, referring to the base space where the
1+1 dimensional theory lives. It is natural to infer therefore that
the 1+1 dimensional model flows in this limit (which is its IR limit)
to a conformal field theory of the heterotic string on $X$, thus
realizing the duality of the M--theory configuration to the heterotic
one as a flow\foot{The issue of realizing duality as a flow from
D--brane world volume theories was discussed a while ago in the final
section of ref.\robme.}. This would be a complicated example of the
work of Motl\motl, and of Banks and Seiberg\banks, and of Dijkgraaf,
Verlinde and Verlinde\dvv. There a ``matrix string theory'' of the
light cone type~IIA string was recovered from Matrix theory as a flow
to an $S_N$ orbifold conformal field theory from a 1+1 dimensional
model arising essentially from a collection of $N$ D1--brane probes'
common world--volume. In that case, with D1--branes in flat space, the
moduli space (and hence the target space of the sigma model) was
$(\IR^8)^N/S_N$ (where $S_N$ is the group of permutations of $N$
objects), which ultimately defined the conformal field theory. (The
uncompactified heterotic matrix theory was studied in
refs.\matrixheterotic, and some heterotic toroidal compactifications
in refs.\refs{\matrixheteroticii,\matrixheteroticiii}.) In this case,
the moduli space of the $N$ D1--brane system needs to be carefully
examined (we have not done it fully here) to discover the precise
nature of the conformal field theory this new matrix string theory
realizes. On general grounds, with everything above as motivation, we
expect of course to define the $E_8{\times}E_8$ heterotic string on
$X{\times}{\cal I}$ in this way.

\subsec{\sl Matrix Theory of M--theory on $K3{\times}{\cal I}$ and its 
Heterotic Limits.}

Next, we come to ask about the definition of M--theory on
$K3{\times}{\cal I}$.  Restricting ourselves to the point we know best
---the special point--- we can anticipate that a six dimensional
theory with $(1,0)$ supersymmetry is involved in defining this model.

At least part of this theory is motivated in terms of
the world volume theory of $N$ D5$^\prime$--branes compactified on
$\widetilde{K3}{\times}S^1_5$, along the lines described in
section~3.3.2. This probe theory should have some interesting
properties, and cannot be the naive super Yang--Mills theory, for the
same reasons as discussed in
refs.\refs{\rozali,\berkoozii,\berkooziii}\ in the case of the
$(0,2)$ situation.

What properties must we require this compactified $(0,1)$ theory to
have?  One of the motivations of this paper is to point out that at
least at a special point in moduli space, {\it the theory must
reproduce the special duality properties of the DMW point:} There
should be dual realizations of the $E_8{\times}E_8$ heterotic string
theory compactified on $K3$ with the (12,12) arrangement of
instantons\foot{A crucial phenomenon we observe here is the fact that
the strong coupling limit of going to M--theory mixes up the role of
the various hypermultiplets in our spectrum in a way which would not
have been consistent with string theory if we had remained at weak
coupling. As pointed out last section, of the 244 hypermultiplets not
involved in removing various $U(1)$'s, 4 of them are closed string
fields, and the rest are 16 copies of 15 (5--9) open string fields,
associated to the 3/2 instantons at each fixed point. By the time we
get to M--theory on $K3{\times}{\cal I}_5$, we have (of course) the
same number of hypermultiplets, but their origin is different. 112
hypermultiplets come from placing 12 $E_8$ instantons on the $K3$ and
breaking the $E_8$ gauge symmetry associated with the end of the
interval completely, and another 112 from the other end of the
interval. The remaining 20 hypermultiplets come from the family of $K3$
deformations. So we see that at each fixed point, a 5--9 field becomes
associated at strong coupling with a deformation of $K3$, while the
other 14 are the 224/16 $E_8$ instanton deformations.}.

This  works as follows: Heterotic/heterotic duality has its
M--theory origins in terms of electromagnetic duality exchanging M2--
and M5--branes, as first suggested in ref.\duff\ and made explicit in
the present paper.  In sections~2.6 and ~2.7, we showed how to reduce
the problem of studying the details of heterotic/heterotic duality in
M--theory to one of studying T--duality in the type~IA model, and a
related T--duality in the type~IB model, where in that context its
relevance was shown in ref.\berkoozi.  So the existence of
heterotic/heterotic duality is guaranteed by the existence of a
$T_{456789}$--duality property which exchanges the two type~IA models
(realizing electric--magnetic duality in M--theory), and hence by a
$T_{6789}$--duality which exchanges the two type~IB models.

This translates therefore in this matrix theory context to the
statement that (at this special point) heterotic/heterotic duality is
ensured by asking that the compactified $(0,1)$ six dimensional theory
has as a basic property a $T_{6789}$--duality symmetry, giving us the
same theory back. (In this limit where we are studying everything
explicitly as orbifolds, such a duality on the base space makes the
usual good sense. Beyond the orbifold limit, it is not clear what to
do.)

This property, as stressed by Seiberg\seiberg, rules out the $(0,1)$
fixed point field theories as candidates, and rules in the ``little''
string theories as possible candidates for our matrix definition. This
is encouraging, as we saw in section~3.3.2 that we identified
excitations in the compactified model which was a particular family of
stringy instantons. We also deduced that part of their definition was
given by the 1+1 dimensional sigma model of earlier subsections.

The 1+1 dimensional sigma model will play two crucial roles in this
situation then. Most naturally, it will flow to a conformal field
theory and define the heterotic matrix theory of the $E_8{\times}E_8$
heterotic string compactified on $X{\times}{\cal I}$, as described in
section~4.2. Alternatively however, there is a limit where the model
flows to a point where gravity decouples, defining a matrix theory of
a ``little $E_8{\times}E_8$ heterotic string'' in six dimensions,
compactified on the base space $\widetilde{K3}{\times}S^1_5$. Such
a dual role fits all of the data and is conceptually satisfying.

In addition to seeing the stringy excitations of the compactified
little string model, we also saw (in section~3.3.2) clues about
exactly what it means to compactify such a theory on $K3{\times}S^1_5$
for the purposes of our definition. We saw that the moduli space of
compactifications of the $N$ D5$^\prime$--brane theory encodes the
moduli space of choices to be made about the arrangements of
D5--branes and D9--branes in the GP models, which is a particular case
of constructing a dual heterotic compactification: This amounts to
choosing the gauge bundle and instantons carefully in the presence of
the $K3$, breaking the $Spin(32)/\IZ_2$ gauge symmetry accordingly, as
described in great detail earlier. From the point of view of the
little string theory, we are making the same choices about breaking a
$Spin(32)/\IZ_2$ {\it global} symmetry. This seems at odds with the
fact that we concluded that we had the $E_8{\times}E_8$ little string
theory earlier, but recall that precisely at this special point in
moduli space, the ``big'' $Spin(32)/\IZ_2$ and $E_8{\times}E_8$
heterotic strings are $T_5$--dual. That $T_5$--duality must be
inherited by the compactified little string theories as well, and is
another requirement for our compactified $(0,1)$ theory.

So in short, we see that the moduli space of compactifications of the
little heterotic string theory is indeed just that which we want, in
order to reproduce the physics of the compactified (big) heterotic
string at the special DMW point.

\newsec{\bf Beyond the Special Point}

As the theory at the special point is defined in terms of $N$
D5--branes of type~IB, we have implicitly made a statement about the
(0,1) theory associated to the $Spin(32)/\IZ_2$ global symmetry and
its compactifications. This is strengthened by the fact that
strong/weak coupling duality takes us directly to a family of $N$
$Spin(32)/\IZ_2$ heterotic fivebranes, making direct contact with the
construction of Seiberg\seiberg. It so happens that the particular
point in moduli space we have studied has an accidental $T_5$--duality
symmetry to a similarly compactified $E_8{\times}E_8$ little string
theory.  Away from the special point however, one expects that it is
the latter which will define the full moduli space of $E_8{\times}E_8$
heterotic string $K3$ compactifications\foot{Some alternative comments
about the matrix theory for the full moduli space of the heterotic
string on $K3$ are made in the conclusions of ref.\matrixheteroticiii. I
am grateful to S. Govindarajan for pointing this out.}, although the
details of the prescription are harder to motivate. This is because we
no longer have the luxury of a complete D--brane construction to help
us build sigma models for study, and we also do not have as much
control over the properties of $K3$ away from the orbifold limits.

However, it is a highly suggestive that such a definition may not be
far from the mark. If correct, we would also gain a definition of the
$SO(32)$ heterotic string on $K3$ with the standard embedding, as this
is the $(16,8)$ $E_8{\times}E_8$ model\sixteeneight.

\newsec{\bf Discussion and Conclusion}

\subsec{\sl Summary}

This paper was divided into two main parts. First, we did a thorough
examination of some of the properties of the heterotic/heterotic
duality of ref.\duff. This was made possible by the fact that it has a
realization as a type~IB orientifold model\ericjoe, as pointed out in
ref.\berkoozi. 

We exploited this fact by turning this orientifold definition into a
type~IA one, and noticed that precisely at the point of interest, we
can take the limit where the model becomes an M--theory
compactification, and from there made contact again with the heterotic
model. Along the way, we sharpened our understanding of the
heterotic/heterotic dual phenomenon and its links to M--theory's
electric--magnetic duality properties.

Such an exploration is interesting in its own right and we could have
stopped there, but we went further to the second half of the
paper. Given that we have a family of orientifold models which
connects so nicely the heterotic string on $K3$ to M--theory, on one
hand, and to type~I theory on the other, it is natural to see whether
we might make some headway in defining what the nature of a Matrix
theory representation of this M--theory compactification might
be. Indeed, if we cannot say anything else about the theory (due to
its admittedly low amount of supersymmetry, and its points of
spacetime curvature which make D--brane matrix definitions
problematic, at least at finite $N$ \douglasooguri) we can at least try
to constrain its properties by requiring it to reproduce some of the
intricate properties of the heterotic/heterotic situation.

This is what we set out to do in the second half of the paper. Along
the way, we derived a number of interesting observations which have
wider application. The 1+1 dimensional probe model of the D1--branes
moving in the orientifold model background was particularly
interesting, generalizing the work of
refs.\refs{\wittenadhm,\douglasmoore,\douglasii}\ to situations where
the gauge bundles of the heterotic compactification are of just the
right type to realize the DMW duality. 

Now holding some definite models with which to make a Matrix theory
proposal, we went on to discuss the properties of the various matrix
models which would define the various parts of the theory. We
completed the proposal with the conclusion that the central role is
played by the ``little heterotic string theories'', where we can (at
the special point) make a precise statement about the nature of the
base space we are to compactify them on, and the identification of the
moduli space of those compactifications with those of the usual,
``big'' heterotic strings. It is natural to suppose that away from
the special point, the compactified (0,1) little string theories
should be used to define the matrix theory representations of the
wider class of heterotic compactifications too, although there is less
evidence to support this conjecture.

\subsec{\sl New Directions}

There is a wealth of interesting issues uncovered here. In particular,
more study of the 1+1 dimensional model presented here should be
carried out. Part of the moduli space theory contains the physics of
$E_8$ instantons on the Eguchi--Hanson space, which is interesting in
its own right. Furthermore, the model can be used to define (by RG
flow) an interesting conformal field theory of the matrix heterotic
string in such a background, and the details of the moduli space needs
to be pinned down in order to identify correctly the properties of the
candidate conformal field theory. Ultimately, deformation of the
conformal field theory should help construct (by analogy with
refs.{\motl,\dvv)) the full string field theory perturbation series
for heterotic strings in that background.

In another limit this 1+1 model will actually flow to the little
heterotic string itself. This is a more careful limit where gravity
decouples, and the resulting string is forced to live in the six
dimensional situation described above, now compactified on
$\widetilde{K3}{\times}S^1_5$. The determination of the moduli space
of the model will be important for this application too. We have only
completed a partial study in this paper.

It is interesting to consider also the case of other orientifold
models. Generically, most such models will not satisfy our criteria of
section~2.3.2 for being able to be taken properly to M--theory\foot{I
am grateful to R. C. Myers for discussions on the interesting issue of
whether there are other ways of controlling the limit of six
dimensional type~IA orientifold models in such a way as to get
M--theory configurations.} There is at least one example of a model
which passes the test. The ``$\IZ_4^A$'' model of ref.\ericmeI\ has an
arrangement of D4--branes which allows complete local cancellation of
dilaton charge, thereby promising an interesting dual realization. A
particularly interesting property of that model is the fact that it
has $n_T{=}4$; there are four extra tensor multiplets in the
spectrum. As pointed out in ref.\ericmeII, there is strong evidence
that it is related to the $(10,10)+4$ M--theory $K3_{\rm
M}{\times}{\cal I}$ compactification, having four M5--branes in the
interior of $\cal I$. (It was also argued\foot{In that paper, similar
statements also were made for the $\IZ_6^A$ orientifold model of
ref.\ericmeI. This has since been shown by E. Gimon\eric\ to be
incorrect. It is not possible to arrange the D5--branes in the way
suggested there in order to cancel the dilaton charges locally, and
therefore more care is needed with this model.}\ in ref.\ericmeII\ to
be a limit of F--theory on a Calabi--Yau manifold with Hodge numbers
(127,7), which has since been shown to be true\ericper.) Such a model
is therefore not completely to be understood as a heterotic string
compactification. Because of its special properties as an orientifold
model however, and in the light of this paper, it might be used as a
test--bed for a matrix theory representation of the $(10,10)+4$
M--theory compactification. Much the same procedures as carried out in
this paper would be relevant to such a study. An obvious guess at the
result would be that the $E_8{\times}E_8$ little string theory should
be compactified on an appropriate dual surface (which is meaningful
because in that model $K3$ is realized as a $\IZ_4$ orbifold of a
torus) with some extra features to give the four extra
fivebranes. Presumably these features involve tuning four of the 24
$E_8$ constituent instantons such that they move out onto the branch
of their moduli space where a tensor arises in exchange for 29
hypermultiplets.

There are many avenues of interest to pursue using the constructions
of this paper as a starting point.  Hopefully such further research
will help to further clarify the properties of Matrix theory and the
little string theories.


\vfill\eject

\noindent
\centerline{\bf Acknowledgments}

\medskip

\noindent
CVJ was supported in part by family, friends and music. Thanks to the
Aspen Center for Physics, where most of this work was carried out, and
the organizers of the Duality workshop (July
13th to August 3rd 1997), for an inspiring meeting. Thanks also to the
organizers of the International Symposium on `New Trends in Subatomic
Physics', (National Taiwan University, August 5th to 13th 1997), for
an interesting and timely gathering, and where most of the rest of
this work was carried out in the remarkably fine surroundings of the
Grand Hotel, Taipei.

CVJ would like to thank Ofer Aharony, Eric Gimon, Ami Hanany, Robert
Myers, Hirosi Ooguri, Eva Silverstein and Edward Witten for
conversations.

Travel support was obtained under Department of Energy grant \#
DE--FC02--91ER75661.

\bigskip
\bigskip

\centerline{\epsfxsize1.0in\epsfbox{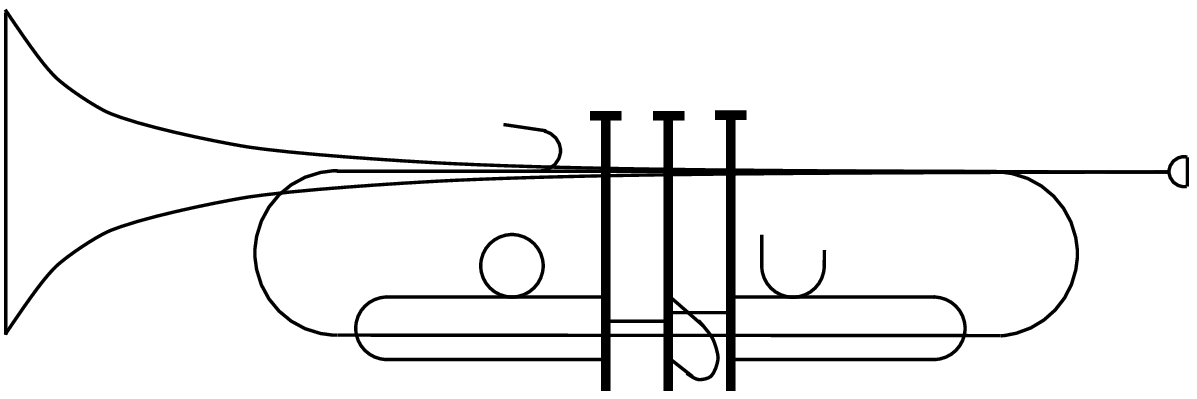}}

\listrefs

\bye